# Fifty Years of ISCA: A data-driven retrospective on key trends

Gaurang Upasani[*], Matthew D. Sinclair[^], Adrian Sampson[+], Parthasarathy Ranganathan[*], David Patterson[*,%], Shaan Shah, Nidhi Parthasarathy, Rutwik Jain[^]

+ Cornell University * Google
% University of California, Berkeley ^ University of Wisconsin-Madison

[gupasani, parthas]@google.com
[msinclair, rnjain]@wisc.edu
asampson@cs.cornell.edu
pattrsn@cs.berkeley.edu
[shaan.shah365,nidhi.parthasarathy]@gmail.com

## 1. Motivation

Computer Architecture, broadly, involves optimizing hardware and software for current and future processing systems. Although there are several other top venues to publish Computer Architecture research, including ASPLOS, HPCA, and MICRO, ISCA (the International Symposium on Computer Architecture) is one of the oldest, longest running, and most prestigious venues for publishing Computer Architecture research. Since 1973, except for 1975, ISCA has been organized annually. Accordingly, this year will be the 50th year of ISCA. Thus, we set out to analyze the past 50 years of ISCA to understand who and what has been driving and innovating computing systems in that timeframe. This analysis is intended to be a celebration of the first 50 years of ISCA. Thus, the scope should be viewed accordingly. Although we took care to practice good data collection and sanitation in our analysis (Section 2), given the long time frame and issues with digital records for early years of the conference, there may be some errors and rounding-off artifacts. Please reach out if you have any corrections and we can update our Arxiv draft to reflect this errata. Finally, while the collected data and analysis highlight several interesting trends, akin to the cautionary comment from the ISCA Hall of Fame website ("*A real Hall of Fame should be determined by impact, not paper count.*"), we want to acknowledge that some of our numbers may only reflect a partial narrative. That said, our exercise still highlights several interesting trends that we think will be insightful to the broader community.

## 2. Methodology

We gathered information on how the community and the mechanics of ISCA evolved over time, how the industry and content of ISCA evolved over time, and trends about the people publishing in ISCA. In total, there have been 2134 papers published at ISCA. Thus, manually analyzing and categorizing each paper would have been time consuming, error prone, and challenging. Instead, we created a scripting ecosystem that leverages DBLP and Google Scholar to gather as much of the information as possible. We chose to use both DBLP and Google Scholar because each had a subset of the desired information in an easily accessible format. For example, Google Scholar makes it easy to count citations of papers, while DBLP makes it easy to perform more historical searches on ISCA publications in a specific year and or by an author. Moreover, the DBLP-based scripts build naturally off the scripts already used to maintain the ISCA Hall of Fame, which itself leverages DBLP and a modified version of the CSRankings [1] scripting infrastructure. Although these corpora are not guaranteed to have the exact same information[1], we tried to keep each experiment self-contained and self-consistent with the data for that experiment. However, we did not try to reconcile data from the sources for various reasons like temporal consistency issues, naming issues (e.g., DBLP and Google Scholar may consider "David Patterson" and "David A. Patterson" to be different people), and schema/formatting differences. Finally, in some situations, especially for ISCA instances that predated the modern Internet, neither DBLP nor Google Scholar had the necessary information. In these situations, we used paper copies of each year's ISCA proceedings (e.g., the general chair's and program chair's welcome notes) as the final authority on information like number of submissions or accepted papers. Here we thank Mark Hill for his hand-me-down collection of ISCA proceedings, which proved invaluable.

Additionally, when binning data across decades, because ISCA did not occur in 1975, we chose to represent the decades as: 1973-1982, 1983-1992, 1993-2002, 2003-2012, and 2013-2022. Note that this means the first decade only contains nine years of publications, while the subsequent decades contain ten years of publications. Moreover, in several cases we chose not to include the upcoming 2023 publications for consistency. Likewise, in some cases (for example, creating word clouds), we selected 1973, 1983, 1993, 2003, 2013, and 2023 as "sample" years to examine trends – examining the yearly trends would have resulted in both significant overlap and significant processing overhead, without significantly adding to the results.

Over the years, ISCA has also seen significant growth and changes in many facets. One of the more interesting growth vectors has been in how other papers were cited and built in. ISCA predates search engines and digital libraries. For example, ACM's digital library [2] started in 1998 and IEEE Xplore started in 2000 [13]. Finding related work in the 1970s and 1980s was a lot of work – you had to go to libraries to read physical papers and to chase references – which reduced the number of references per paper. Moreover, ISCA (and other conferences like ASPLOS) changed policies on paper length for references in ~2015, hoping to increase the number of citations to computer architecture papers. The addition of Arxiv, newer conferences like HPCA, and existing conferences like ASPLOS becoming annual instead of biannual, also collectively increase the number of Computer Architecture papers per year. As a result, there are roughly 5x more Computer Architecture papers published in top venues per year in 2023 than there were in 1990. Consequently, publishing more papers per year and encouraging those papers to cite more related work has significantly increased the number of citations for more recent ISCA papers compared to ISCA papers from the 1970s and 1980s. For example, the 27 papers of the 1978 ISCA averaged 12 references (316 total), while 40 years later ISCA had 63 papers that averaged 58 citations or 5x more (3636 total). The ISCA 2018 total was ~12x more than 1978. Thus, there might be 25x more citations from refereed Computer Architecture papers published in 2023 than from those in 1990. We also find that newer papers are more likely to cite other, more recent papers than older papers. Likewise, the number of authors per ISCA paper has been increasing over the years [8][12]. This evolution means that more recent ISCA papers are more likely to be highly cited, and since there are often more authors on those papers it is more likely they will get cited – which impacts and biases metrics like examining citation count across all ISCA instances.

# 3. Trends from ISCA's Data

## 3.1 ISCA Community and Mechanics: Growth Trends

[1] In the process of gathering this data, we submitted a number of bug fixes to DBLP to resolve issues with ISCA publications. Interestingly, many of these issues came from ISCA 1992, potentially because ISCA accepted full-length papers and extended abstracts that year.

Figure 1 shows the number of papers submitted to ISCA, the number of papers accepted at ISCA, and ISCA's acceptance rate, across the first 50 years of ISCA. Moreover, the green, red, and blue lines show the rough trends for each of these metrics, respectively

**ISCA Paper Submissions**: The growing number of submissions across the years reflect the growth of the Computer Architecture community: ISCA 1973 had roughly 70 paper submissions, while ISCA 2023 had around 400 – a 6X increase in paper submissions. However, this growth has not always been linear. For example, ISCA 1988 has a significant spike in the number of paper submissions relative to the previous and subsequent years. More broadly, ISCA in the late 1980s and early 1990s sometimes saw the number of paper submissions decrease year-over-year. In the case of the big increase in submissions for ISCA 1988, we believe this can be explained by ISCA occurring in Hawaii for the first time. However, it is less clear why other years in this timeframe saw fluctuations relative to the prior year. One possible explanation is that ISCA was attempting to keep to a single track. While this seems like it would impact acceptance rate (discussed further below) more than paper submissions, perhaps this also had a chilling effect on the number of submissions as well. This was also the time period when a few other architecture conferences were ramped up, so there might have been a split in where papers were submitted leading to some fluctuations. Nevertheless, additional analysis as to why paper submissions decreased in this time frame is warranted.

**Accepted ISCA Papers & Acceptance Rate**: Although the paper submissions have significantly increased in the last 50 years, the number of accepted papers (blue in Figure 1) has grown much more slowly over the first 50 years of ISCA: from ISCA 1973 to ISCA 2023 the number of accepted papers has only grown by 2X, despite the 6X increase in paper submissions. Accordingly, the ISCA acceptance rate (red in Figure 1) has steadily declined from ~50% (ISCA 1973) to ~20% (ISCA 2023). We believe this highlights the increasing selectiveness of ISCA as the prominence of the conference grew.

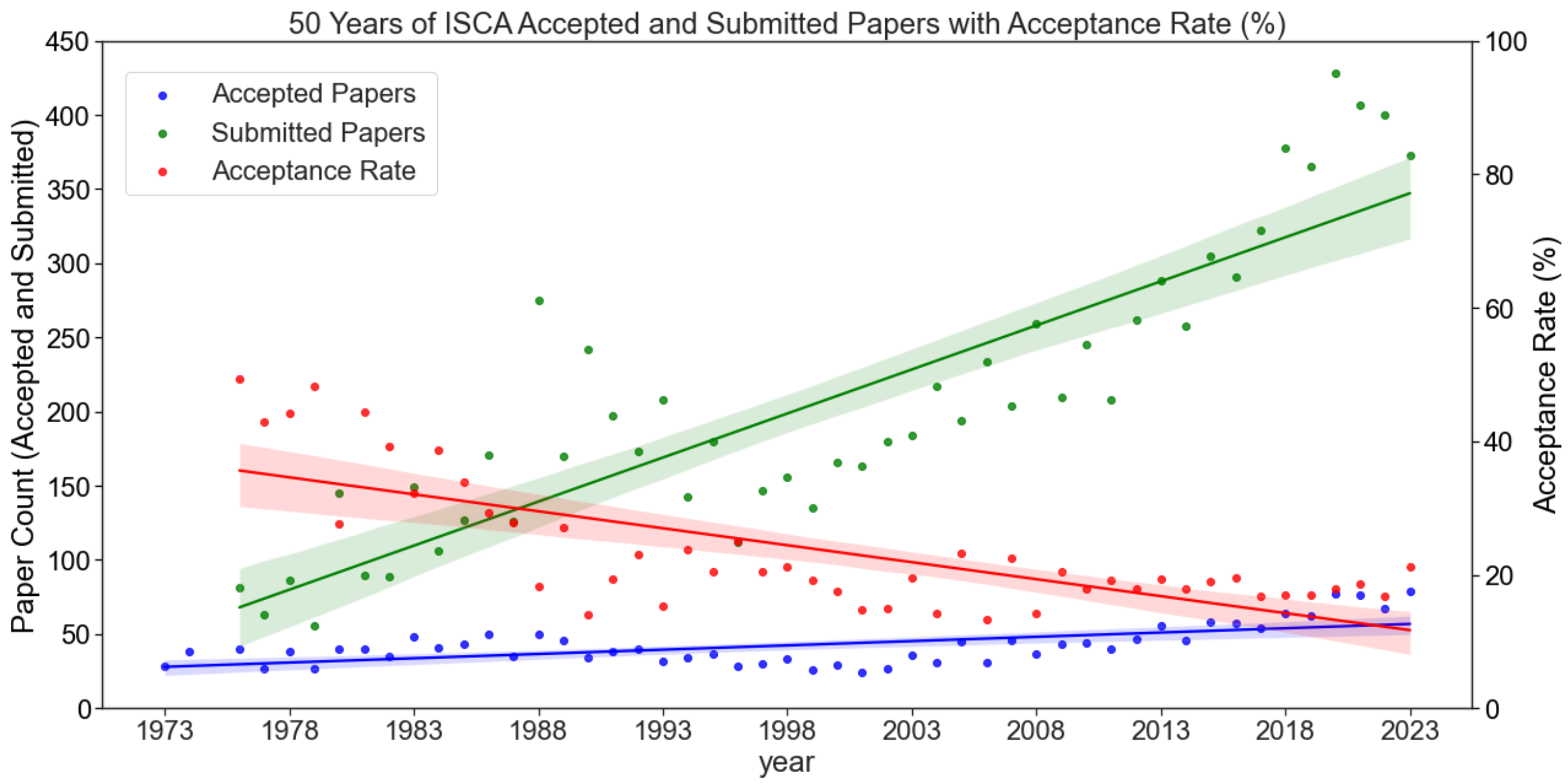

**Figure 1: How the number of submissions, accepted papers, and acceptance rate varied for the first 50 years of ISCA.**

**Number of Authors Per Paper:** Figure 2A-B shows how the mean number of authors per accepted ISCA paper varied across the first 50 years. Prior work in the late 1990s showed that the average number of authors per paper was steadily increasing at ISCA [12]. Figure 2A-B shows that this trend has largely continued to the modern day. While early years of ISCA had a number of papers with single, or very few, authors (e.g., a mean of 2.1 authors in 1973), the data shows that these papers are becoming increasingly rare. As highlighted in prior work [12], we believe this highlights the increasing difficulty and challenges in building and researching increasingly large, more complex systems and the accompanying methodology (e.g., simulators, FPGA platforms, or real systems) required to perform the research. However, interestingly the trend is not completely linear: for example 2007 - 2010 saw the mean number of authors decrease from 4.7 to 3.7. It is unclear why this drop happened though, as there were papers such as the Anton [27] and Corona [28] papers with large author counts in this timeframe. Thus, we suspect there is perhaps some noise in the cadence of paper acceptances here. There is also a large spike in the mean number of authors between 2019 (5.0) and 2020 (6.6). We believe this spike correlates with the introduction of an industry track [11], as well as several industry papers in recent years such as AMD's Exascale paper [20], Google's TPU papers [24][25] and Groq's TSP papers [21][22]. Each of these papers had at least 22 authors, and in general papers on building large-scale projects such as these tend to have many authors, causing the mean to increase significantly as these papers became accepted more frequently at ISCA. However, even in other papers besides these, the number of authors appears to be steadily increasing.

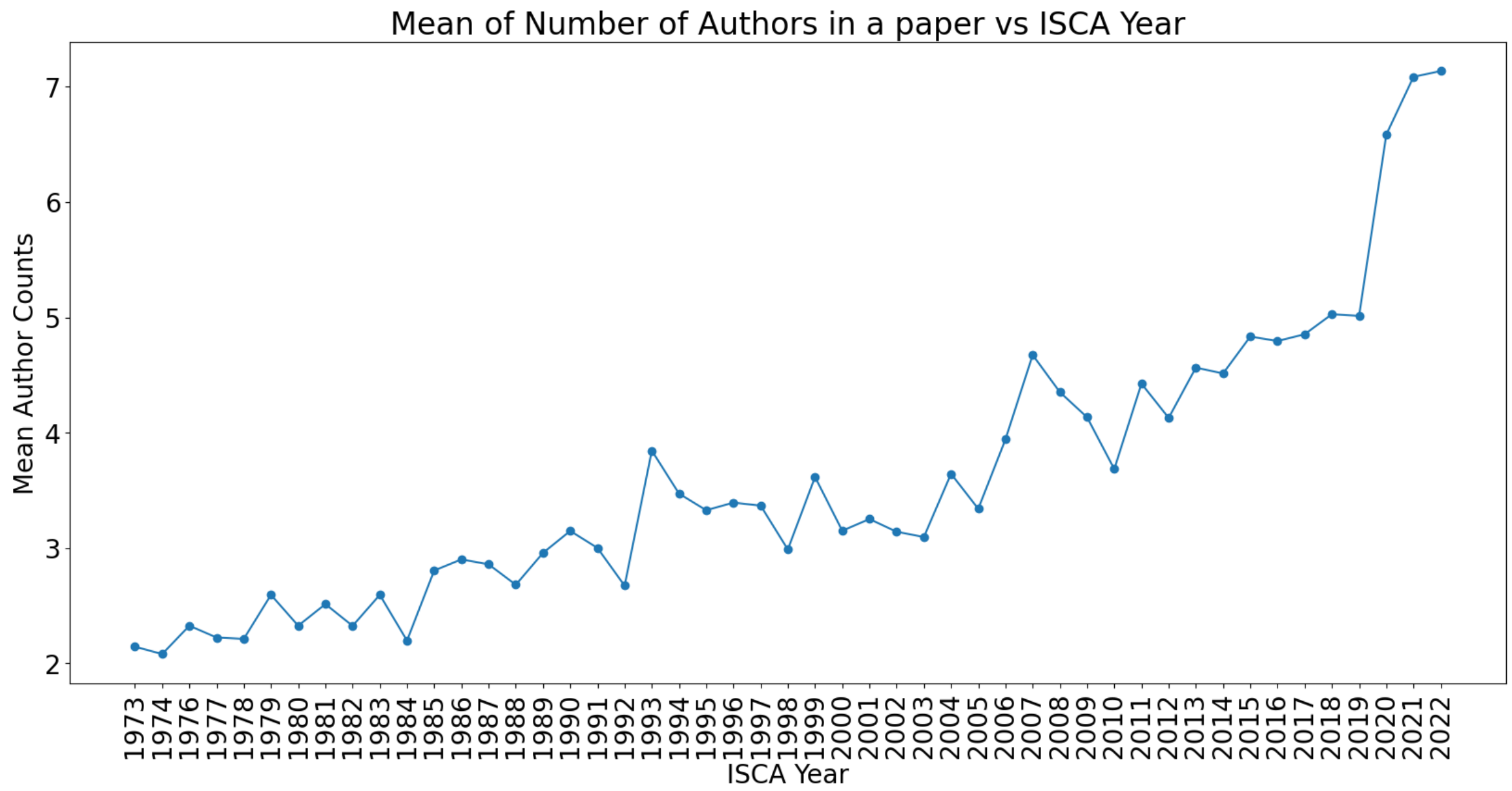

**Figure 2A: How the mean number of authors per accepted paper varied for the first 50 years of ISCA.**

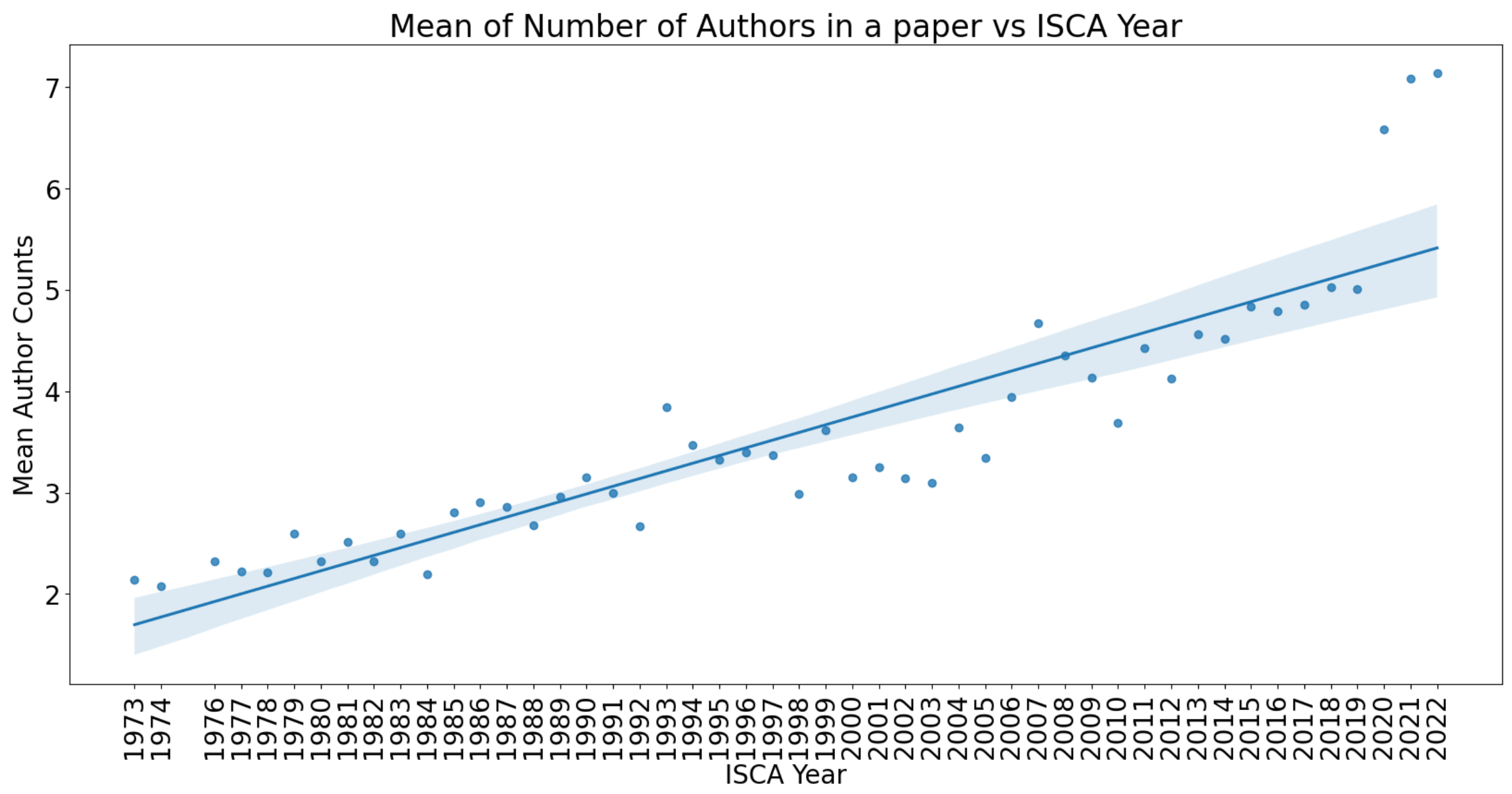

**Figure 2B: Using a trendline to plot the growth of the mean number of authors per accepted paper for the first 50 years of ISCA.**

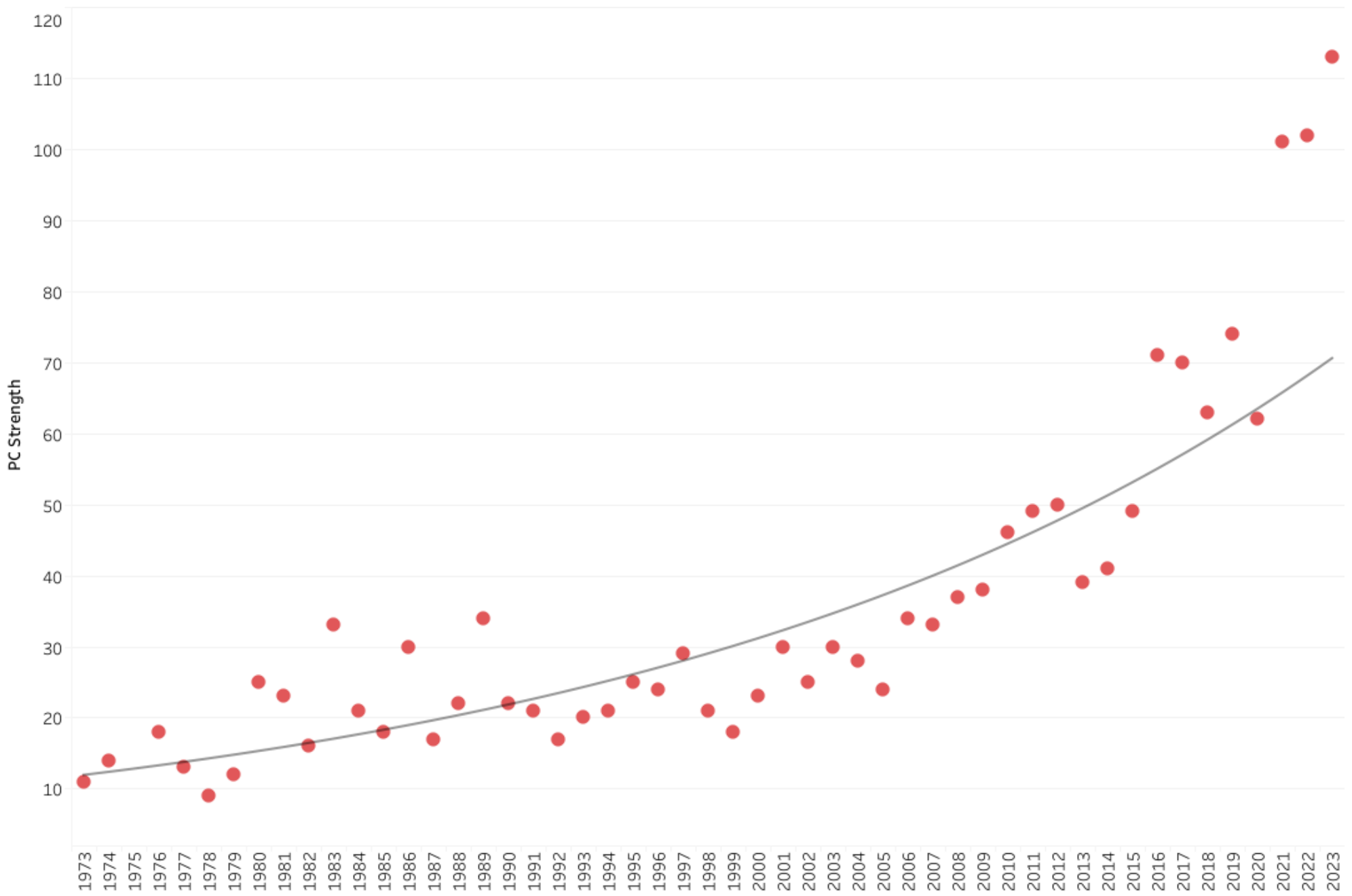

**Figure 3: How ISCA's Program Committee size varied for the first 50 years of ISCA.**

**Program Committee (PC):** Given the large increase in paper submissions, the ISCA main program committee (i.e., not including external reviewers or referees) has needed to grow. As shown in Figure 3, while ISCA 1973 only had a PC of 11 members, ISCA 2023 had a PC of 113 members – a 10X increase in the number of PC members across the last 50 years! In fact, at last year's ISCA program committee meeting, the number of people in the "conflict zoom room" was sometimes larger than the size of the original ISCA PC from 1973! More generally, the number of people in the PC initially tracked the exponential growth in paper submissions. However, in recent years the PC growth rate has started to exceed the growth rate of paper submissions.

**ISCA Geographical Location:** Figure 4A-B present the geographic locations ISCA has been held at in the first 50 years. Perhaps unsurprisingly given that SIGARCH/TCCA had a rule about hosting 3 out of every 4 ISCA in North America, the majority of ISCAs have been held in the US. Within the US, ISCA has been distributed fairly evenly across the east and west coasts with a strong central presence as well. Internationally, especially in recent years, ISCA has increasingly been hosted in a number of countries in Asia, Europe and North America, including Canada, China, France, Israel, Japan, and South Korea. This continued geo-diversity of where ISCA is held will continue next year, as ISCA will be held in Buenos Aires, Argentina - the first time ISCA will be held in South America! More broadly, the increased geo-diversity is both a reflection of the increased internationalization of our community and the growing diversity in authors and paper submissions.

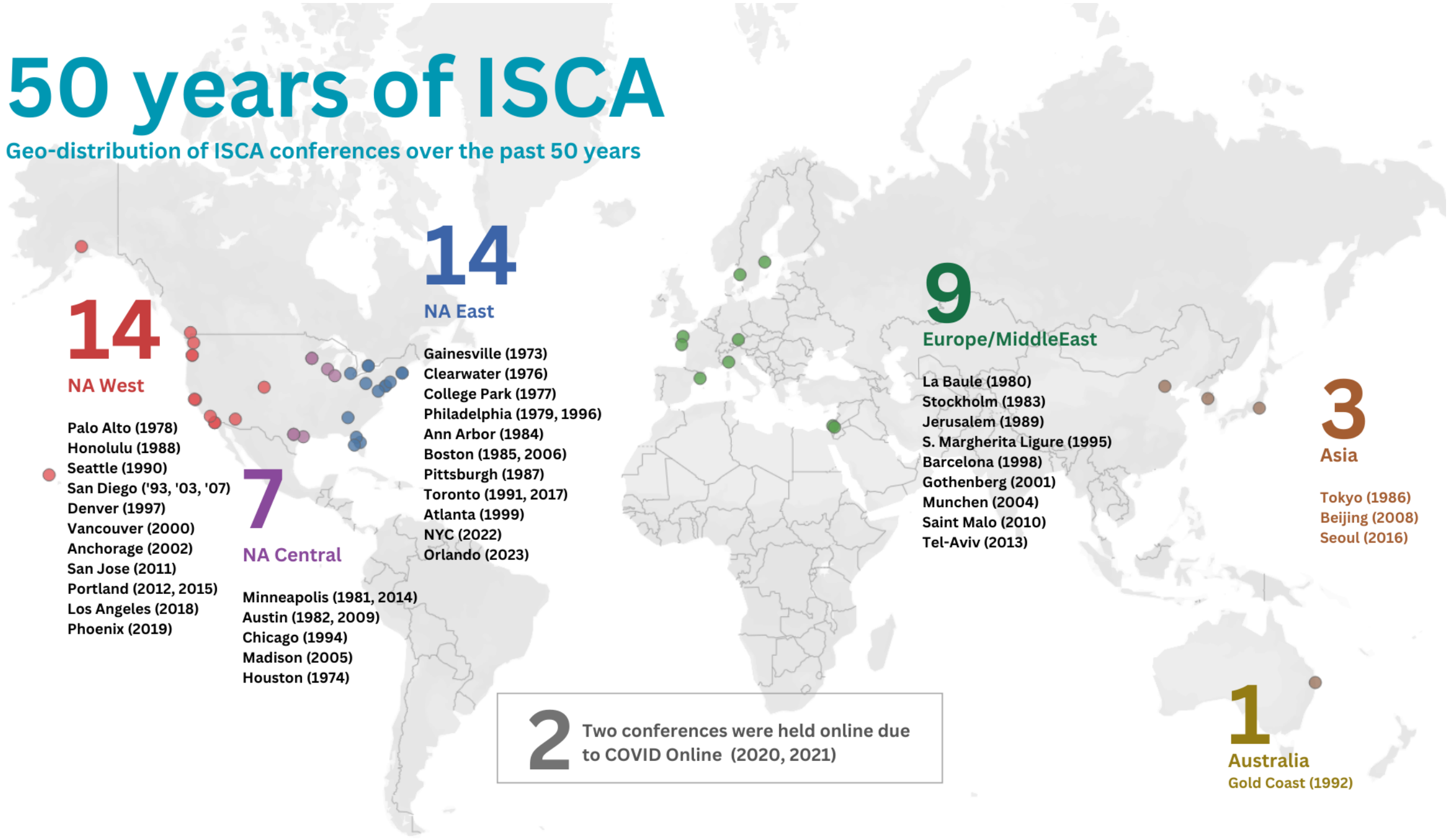

**Figure 4A: Geographical distribution of where ISCA conferences have been held in the past 50 years.**

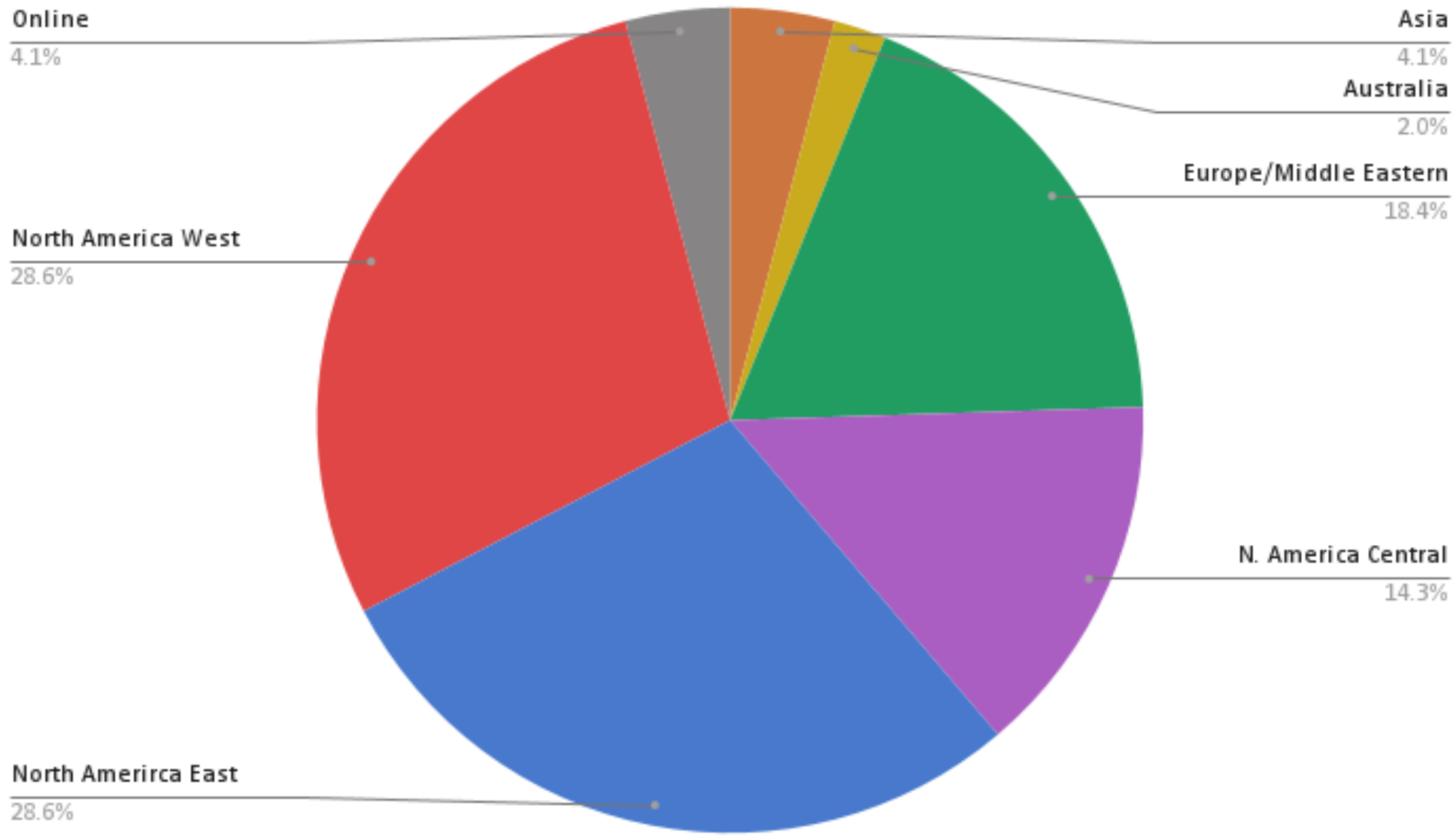

**Figure 4B: Percentage of ISCA conferences held across different geographies**

## 3.2 ISCA Industry and Content Growth Trends

### 3.2.1 Per-Decade Word Clouds

Examining how the keywords in abstracts varied across the decades provides additional insight into how the ISCA's focus and content evolved over the past 50 years. Figure 5A-F shows word clouds of the most common words in abstracts from the ISCA 1973, ISCA 1983, ISCA 1993, ISCA 2003, ISCA 2013, and ISCA 2023. The word clouds show the unique words across all abstracts in each year, after filtering for common words like "the". The number of unique words varies by year and ranges from 164 to 286 words. These word clouds introduce several interesting trends. ISCA papers in the initial decade focused on designing hardware and systems (often for minicomputers), including specialized designs for avionics and meteorology. To a lesser extent, other topics like fault tolerance also appeared prominently in the first year of ISCA – hinting at the growing importance of this field as transistors became smaller and designs more complex. In subsequent decades, parallelism at various levels was a common topic in ISCA 1983, 1993, and 2003. This includes fault-aware computing, multiprocessors, synchronization, memory, and caches, as well as instruction-level parallelism and other techniques to increase parallelism within a given processor. However, as Dennard's Scaling faded, architects increasingly turned further towards parallelism and virtualization, initially continuing prior work on multiprocessors and multi-core CPUs (ISCA 2003, 2013) and later turning more and more to GPUs and other specialized accelerators (ISCA 2013, 2023). Unsurprisingly, topics like accelerators (e.g., machine learning, robotics), quantum computing, and security appear much more prominently in ISCA 2023 relative to prior years, signifying the importance of these workloads in driving the requirements of future systems, as well as the challenges with continuing to scale multi-core CPU performance due to Dark Silicon [18], the end of Dennard's Scaling, and the slowing of Moore's Law.

Figure 6 shows the cumulative distribution for the total number of words for each of the years. The x-axis represents the number of unique words in the word cloud, and the y-axis represents the cumulative total percentage. The data shows the vocabulary in ISCA papers has gotten richer, suggesting increased topic diversity (and/or writing style changes).

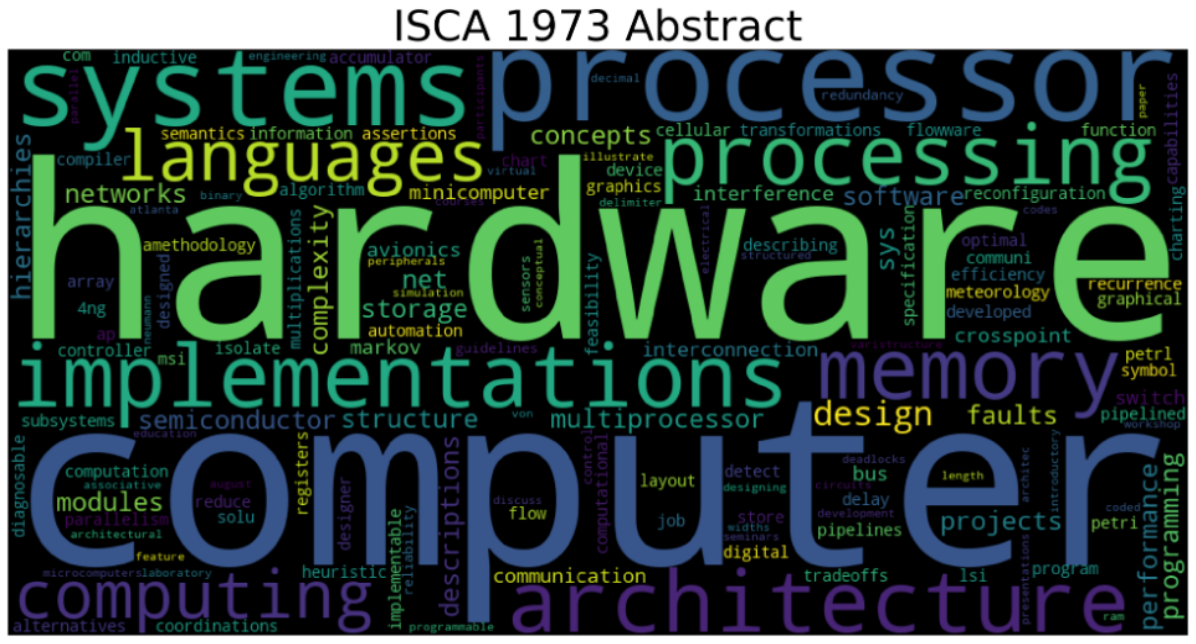

**Figure 5A: Word Cloud of Most Common Words in Abstracts for ISCA 1973.**

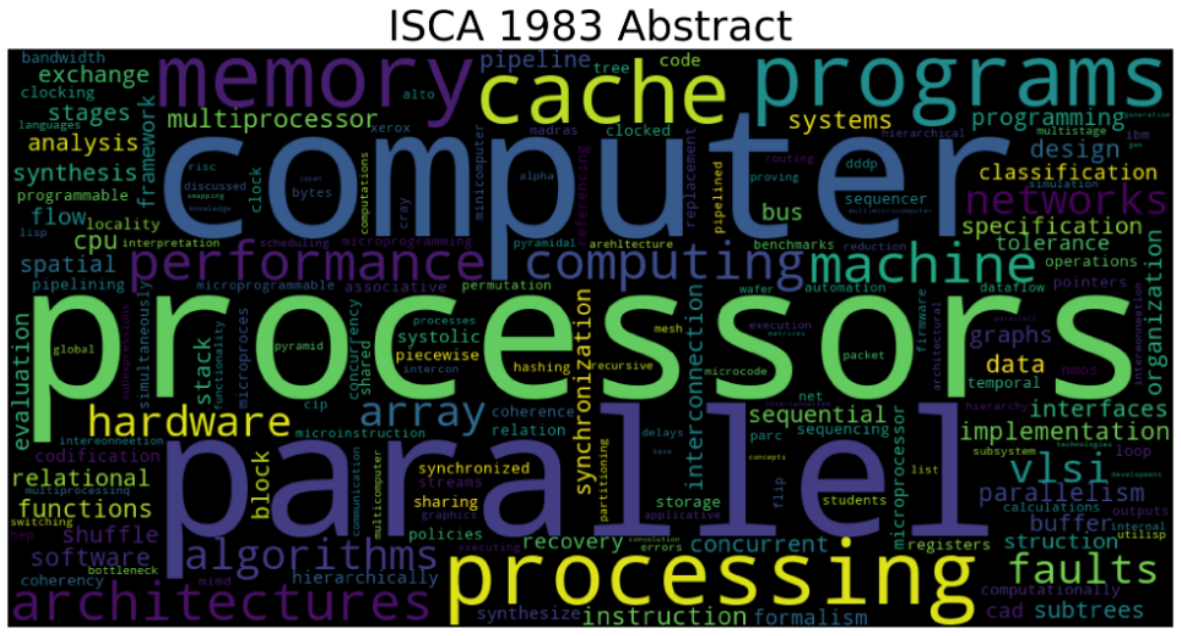

**Figure 5B: Word Cloud of Most Common Words in Abstracts for ISCA 1983.**

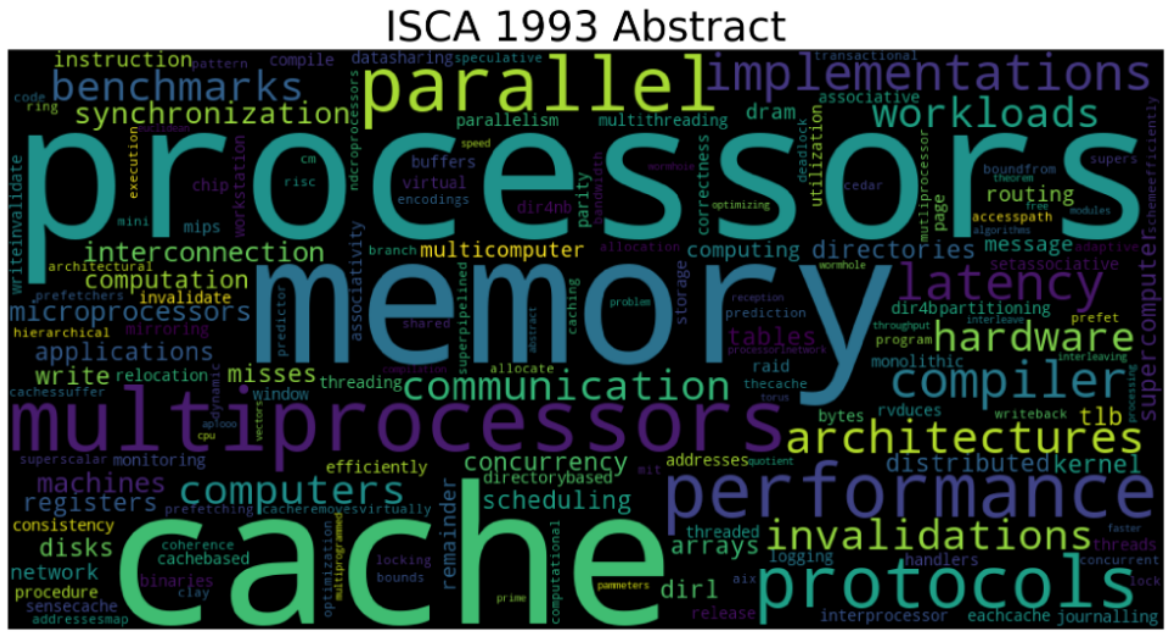

**Figure 5C: Word Cloud of Most Common Words in Abstracts for ISCA 1993.**

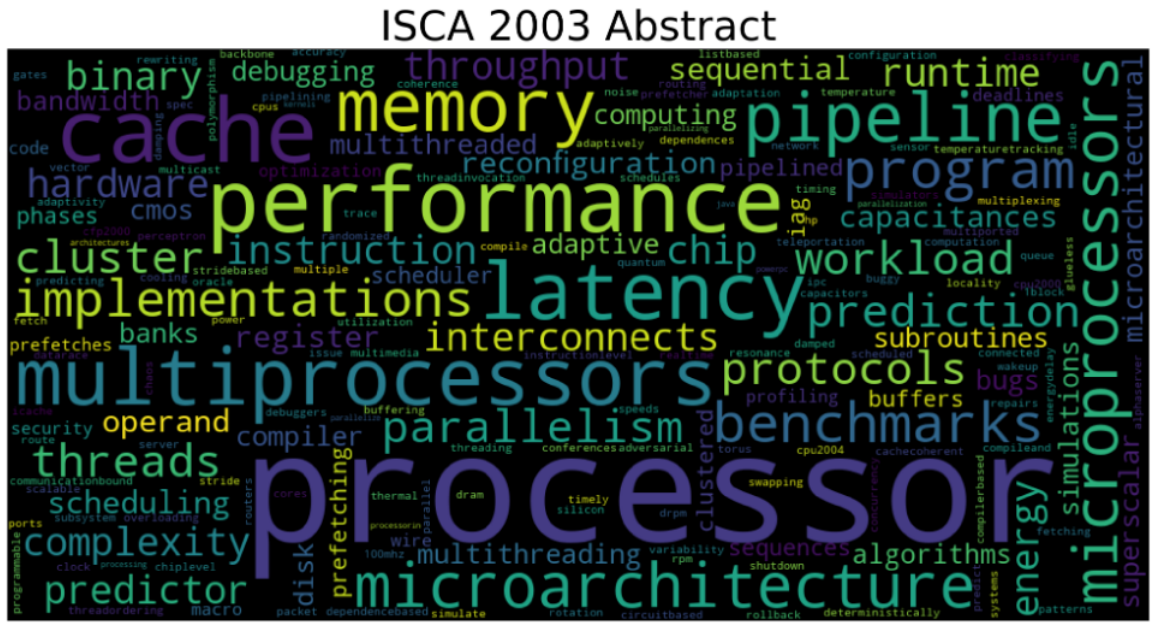

**Figure 5D: Word Cloud of Most Common Words in Abstracts for ISCA 2003.**

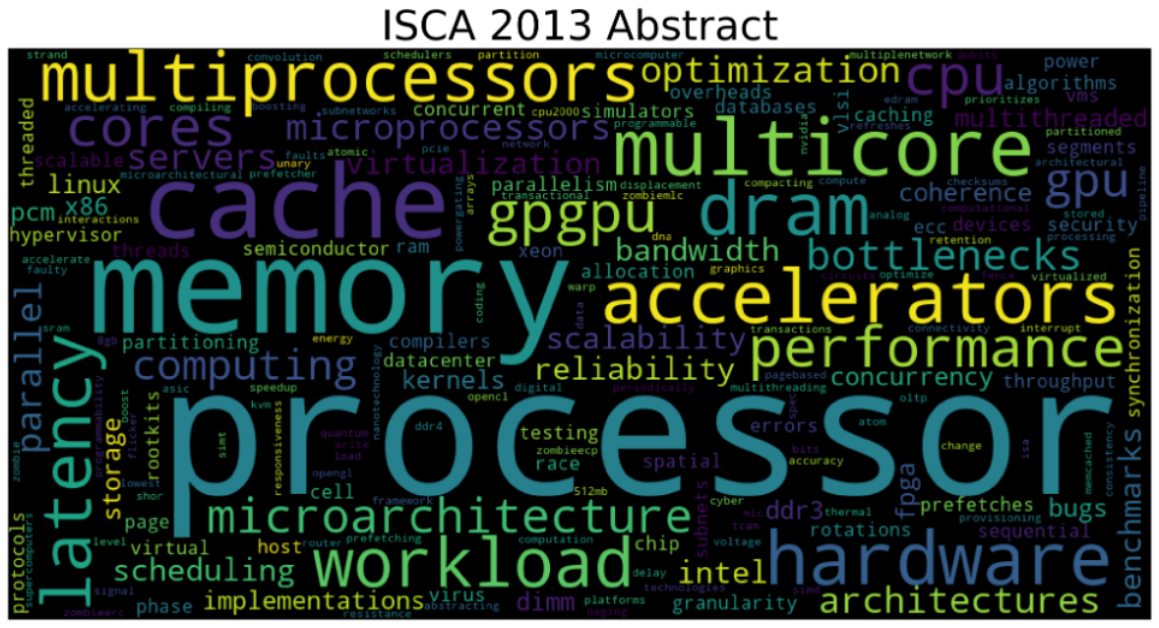

**Figure 5E: Word Cloud of Most Common Words in Abstracts for ISCA 2013.**

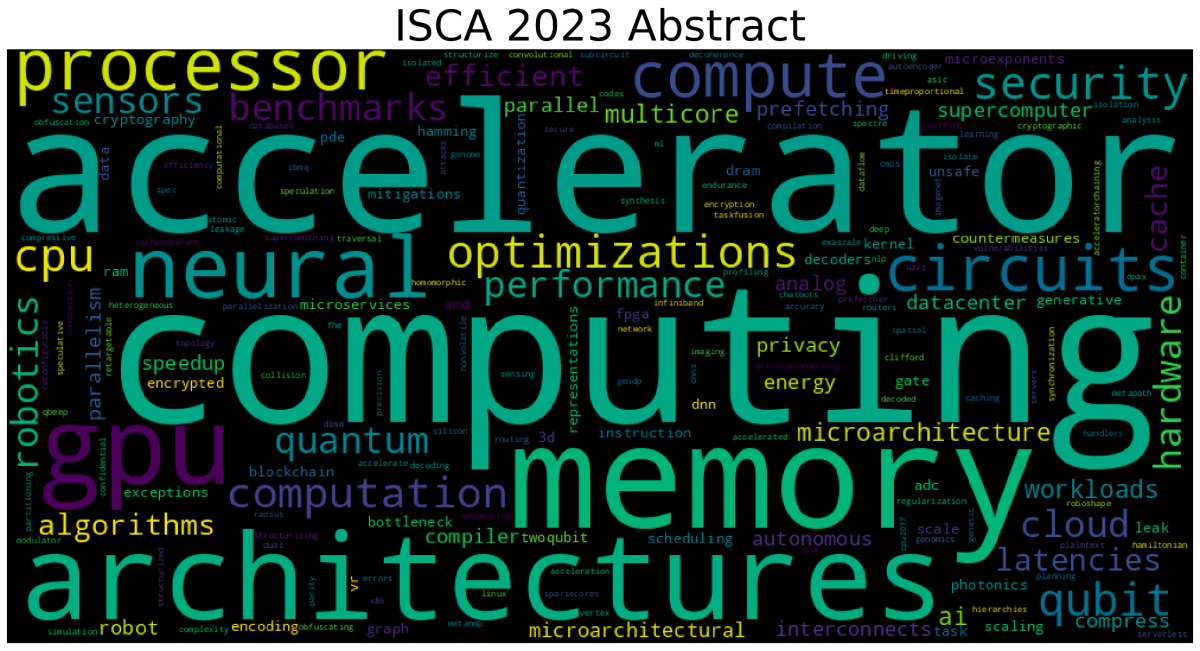

**Figure 5F: Word Cloud of Most Common Words in Abstracts for ISCA 2023.**

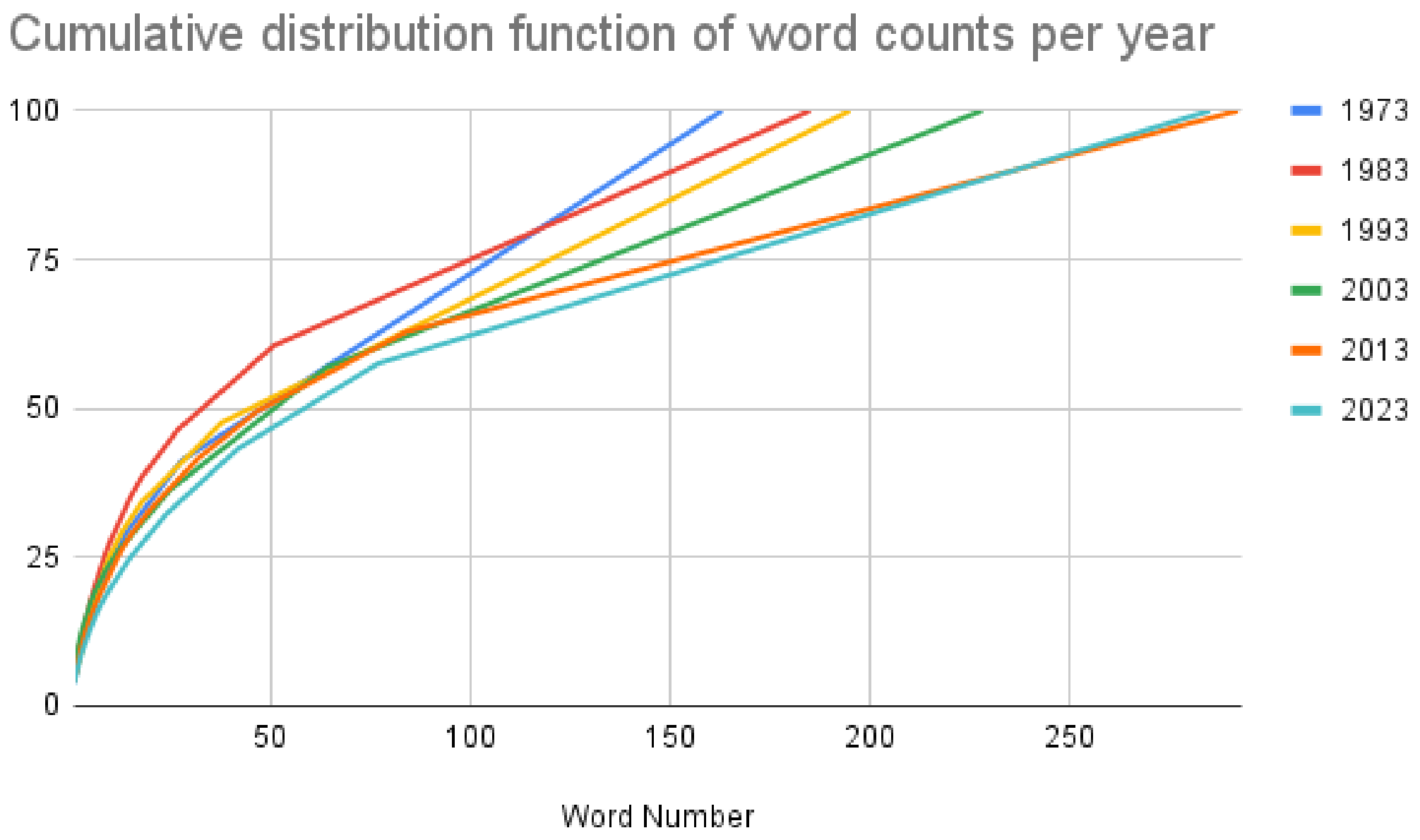

**Figure 6: Cumulative distribution function of word counts per year**

### 3.2.2 Top-Cited Paper Per Year

Table 1 shows the top cited paper (per Google Scholar) for each year ISCA occurred, with the highest cited paper per decade highlighted in green. Figures 7 and 8 further expand on this by using the Type and Topic fields (using the same methodology discussed in Section 3.2.3) to analyze these papers by type and topic. Examining these top-cited papers per year introduces some interesting trends in how ISCA's content and topics evolved over the past 50 years, some of which reinforce our takeaways from the word clouds in Section 3.2.1:

**Table 1: Top Cited Paper For Each ISCA from 1973-2022. The top cited paper from each decade is highlighted in green.**

| Year | Paper | Citations | Type | Topic |
|---|---|---|---|---|
| **Decade 1 (1973-1982)** | | | | |
| 1973 | Banyan Networks for Partitioning Multiprocessor Systems | 937 | Micro | Interconnect |
| 1974 | A Preliminary Architecture for a Basic Data Flow Processor | 854 | Arch | Parallelism |
| 1976 | Improving the Throughput of a Pipeline by Insertion of Delays | 126 | Micro | Pipelining |
| 1977 | A Large Scale, Homogeneous, Fully Distributed Parallel Machine, I | 747 | Micro | Parallelism |
| 1978 | DIRECT - A Multiprocessor Organization for Supporting Relational Database Management Systems | 364 | Arch | Parallelism |
| 1979 | Processor-Memory Interconnections for Multiprocessors | 183 | Arch | Interconnect |
| 1980 | A Comparison Connection Assignment for Diagnosis of Multiprocessor Systems | 711 | Arch | Parallelism |
| 1981 | A Study of Branch Prediction Strategies | 1203 | Micro | Parallelism |
| 1982 | Decoupled access/execute computer architectures | 373 | Micro | Pipelining |
| **Decade 2 (1983-1992)** | | | | |
| 1983 | Very Long Instruction Word Architectures and the ELI-512 | 854 | Arch | Parallelism |
| 1984 | A Low-Overhead Coherence Solution for Multiprocessors with Private Cache Memories | 800 | Micro | Consistency/ Coherence |
| 1985 | Implementing A Cache Consistency Protocol | 408 | Micro | Consistency/ Coherence |
| 1986 | Memory Access Buffering in Multiprocessors | 742 | Micro | Consistency/ Coherence |
| 1987 | Checkpoint Repair for Out-of-order Execution Machines | 342 | Arch | Parallelism |
| 1988 | An Evaluation of Directory Schemes for Cache Coherence | 826 | Micro | Consistency/ Coherence |
| 1989 | Can Dataflow Subsume von Neumann Computing? | 391 | Micro | Parallelism |
| 1990 | Improving Direct-Mapped Cache Performance by the Addition of a Small Fully-Associative Cache and Prefetch Buffers | 2247 | Micro | Cache |
| 1991 | IMPACT: An Architectural Framework for Multiple-Instruction-Issue Processors | 503 | Micro | Parallelism |
| 1992 | Active Messages: A Mechanism for Integrated Communication and Computation | 2489 | Micro | Parallelism |
| **Decade 3 (1993-2002)** | | | | |
| 1993 | Transactional Memory: Architectural Support for Lock-Free Data Structures | 3390 | Micro | Parallelism |
| 1994 | The Stanford FLASH Multiprocessor | 1052 | Arch | Parallelism |
| 1995 | The SPLASH-2 Programs: Characterization and Methodological Considerations | 5361 | Tools | Benchmark |
| 1996 | Exploiting Choice: Instruction Fetch and Issue on an Implementable Simultaneous Multithreading Processor | 1210 | Micro | Parallelism |
| 1997 | Complexity-Effective Superscalar Processors | 1285 | Micro | Parallelism |
| 1998 | Pipeline Gating: Speculation Control for Energy Reduction | 633 | Micro | Power |
| 1999 | PipeRench: A Coprocessor for Streaming multimedia Acceleration | 707 | Tools | Tools |
| 2000 | Wattch: a framework for architectural-level power analysis and optimizations | 3840 | Tools | Tools |
| 2001 | Cache decay: exploiting generational behavior to reduce cache leakage power | 1006 | Micro | Power |
| 2002 | Drowsy Caches: Simple Techniques for Reducing Leakage Power | 1221 | Micro | Power |
| **Decade 4 (2003-2012)** | | | | |
| 2003 | Temperature-Aware Microarchitecture | 1644 | Micro | Power |
| 2004 | Transactional Memory Coherence and Consistency | 1027 | Micro | Consistency/ Coherence |
| 2005 | Interconnections in Multi-Core Architectures: Understanding Mechanisms, Overheads and Scaling | 646 | Micro | Interconnect |
| 2006 | Techniques for Multicore Thermal Management: Classification and New Exploration | 678 | Arch | Power |
| 2007 | Power provisioning for a warehouse-sized computer | 2625 | Micro | Power |
| 2008 | Technology-Driven, Highly-Scalable Dragonfly Topology | 967 | Micro | Interconnect |
| 2009 | Architecting phase change memory as a scalable DRAM alternative | 1806 | Micro | NVRAM |
| 2010 | Debunking the 100X GPU vs. CPU myth: an evaluation of throughput computing on CPU and GPU | 1167 | Tools | Tools |
| 2011 | Dark silicon and the end of multicore scaling | 2405 | Micro | Parallelism |
| 2012 | RAIDR: Retention-aware intelligent DRAM refresh | 617 | Micro | DRAM |
| **Decade 5 (2013-2022)** | | | | |
| 2013 | GPUWattch: enabling energy optimizations in GPGPUs | 710 | Tools | Tools |
| 2014 | A reconfigurable fabric for accelerating large-scale datacenter services | 1406 | Micro | Interconnect |
| 2015 | ShiDianNao: shifting vision processing closer to the sensor | 1081 | Arch | Machine Learning |
| 2016 | EIE: Efficient Inference Engine on Compressed Deep Neural Network | 2727 | Arch | Machine Learning |
| 2017 | In-Datacenter Performance Analysis of a Tensor Processing Unit | 4307 | Arch | Machine Learning |
| 2018 | A Configurable Cloud-Scale DNN Processor for Real-Time AI | 532 | Arch | Machine Learning |
| 2019 | FloatPIM: in-memory acceleration of deep neural network training with high precision | 182 | Arch | Machine Learning |
| 2020 | MLPerf Inference Benchmark | 307 | Tools | Tools |
| 2021 | Ten Lessons From Three Generations Shaped Google's TPUv4i : Industrial Product | 143 | Arch | Machine Learning |
| 2022 | BTS: an accelerator for bootstrappable fully homomorphic encryption | 31 | Arch | Security |
| 2022 | CraterLake: a hardware accelerator for efficient unbounded computation on encrypted data | 31 | Arch | Security |

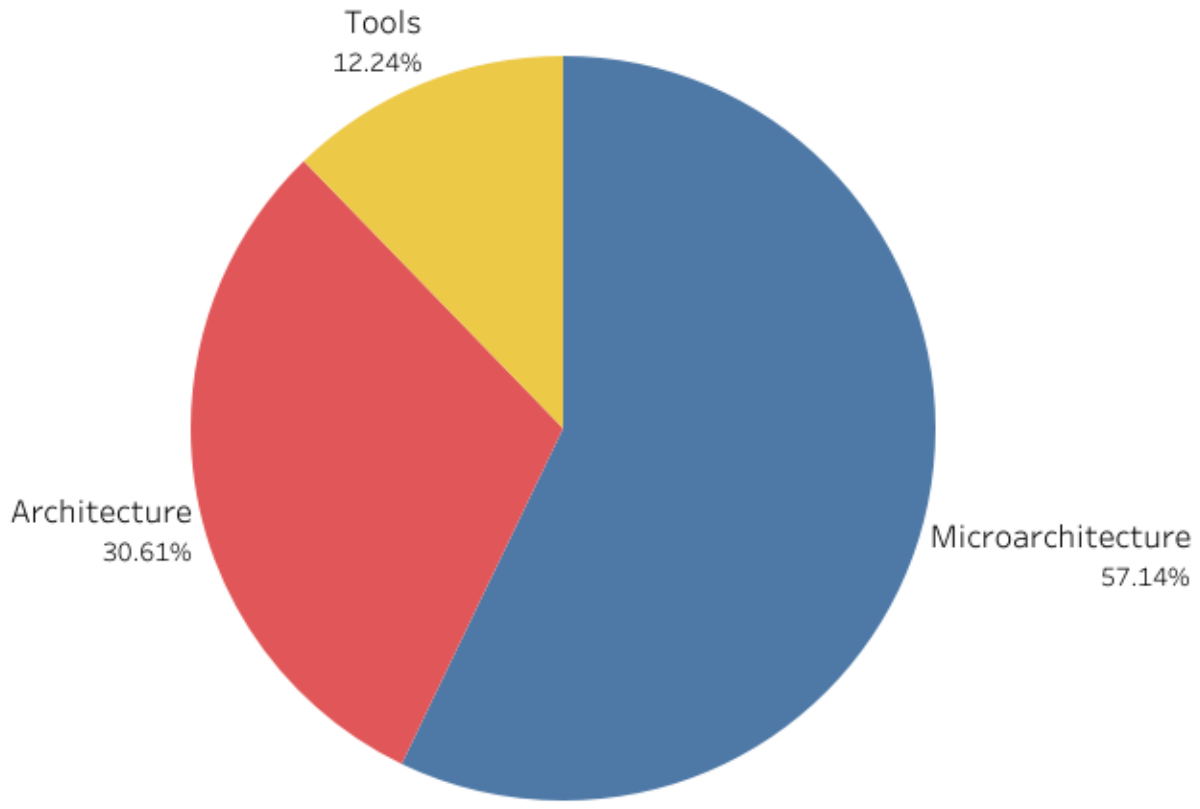

**Figure 7: Per-Year most cited ISCA papers broken down by type.**

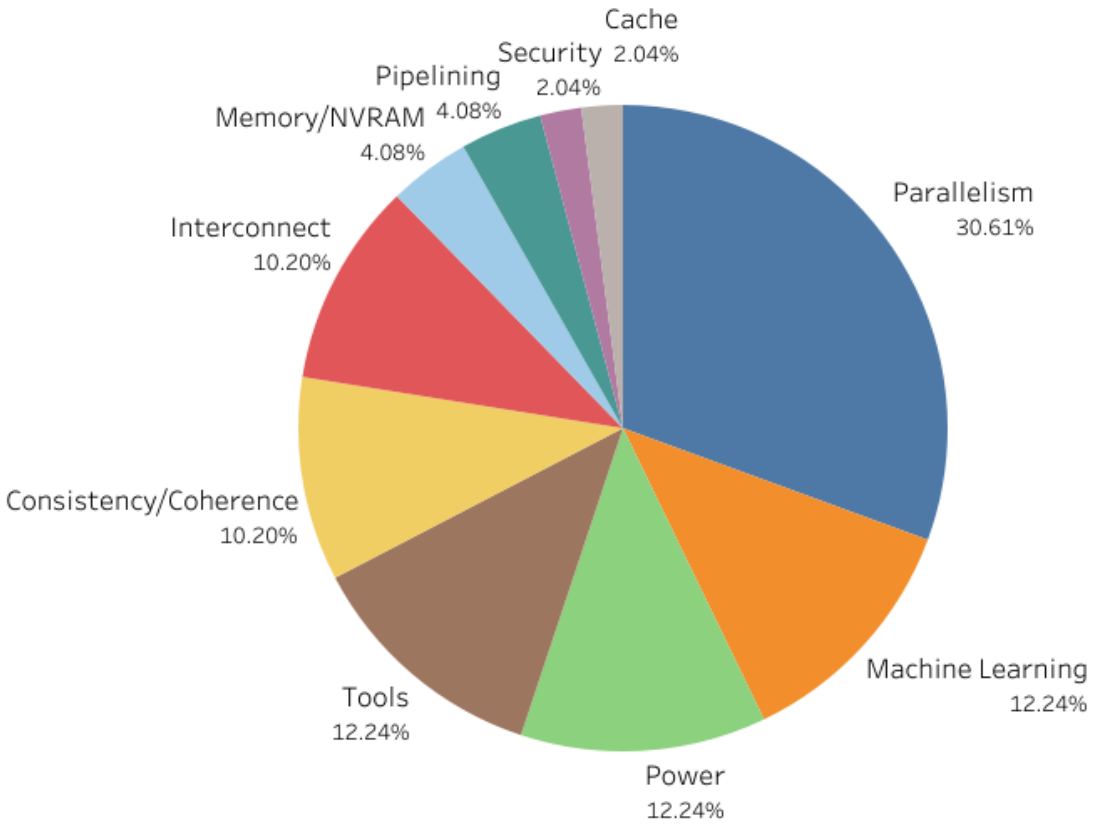

**Figure 8: Per-Year most cited ISCA papers broken down by topic.**

**Decade 1 (1973-1982)**: Multiprocessor papers were on the rise in ISCA's opening decade, as noted by Mark Hill [3]. The papers spanned several topics – networks/interconnects, fault diagnosis, and even relational database systems – all for multiprocessor systems or using multiprocessors. However, the top-cited ISCA paper of the decade is James Smith's study on branch prediction strategies [26]. Although less highly cited, ISCA's first decade also saw papers on other topics such as computer architecture education (e.g., Jonathan Allen's ISCA 1976 paper [19]). While today, these are more likely to appear in specialized conferences, in the 70s, these were still nascent, and their presence in the main program demonstrated ISCA's focus on innovation across a variety of domains.

**Decade 2 (1983-1992)**: VLIW and multiple issue processors add on to multiprocessor papers in ISCA's second decade. Thanks to all this parallelism, cache coherence and consistency become important: papers on these topics were the top cited papers in 1984, 1985, and 1988. Additionally, memory and cache architectures start to appear. For example, the Jouppi paper on victim caches and prefetch buffers [30], for example, is often still required reading for graduate Computer Architecture courses (perhaps unsurprising given that it is 10th on the list of most cited ISCA papers of all time, Table 2).

**Decade 3 (1993-2002)**: 1996 introduced the SMT paper [29] (again, often required reading for graduate Computer Architecture courses). While the top-cited themes include superscalar processors and transactional memory, the highlight of the decade is the focus on power and thermal management. After decades of technology and power scaling, this flurry of activity suggested a leading indicator of the slowing of Dennard scaling that happened in the early 2000s. For example, the 1998 paper on pipeline gating [31] and back-to-back top cited papers from 2000 to 2003 all talk about techniques to reduce power and/or model it.

**Decade 4 (2003-2012)**: ISCA's fourth decade witnessed the boom and, to some extent, the bust in the excitement around multi-core architectures. Almost all top cited papers are either on multicore architectures or interconnects for multicore architectures. Power and thermal considerations carry over from the last decade. Towards the end of the decade, the Dark Silicon paper [18] highlighted major challenges with multicore scaling, which is power limited.

**Decade 5 (2013-2022)**: The natural next step to tackle dark silicon is specialization, and the last decade clearly demonstrates this: ISCA's fifth decade starts off with a top-cited paper on GPU power modeling and reconfigurable architecture, but subsequently Machine Learning dominates the rest of the list. The top-cited papers from 2015 to 2021 are all ML Architecture papers, while ISCA 2022's top cited papers (thus far) are on accelerators for homomorphic encryption and encrypted data.

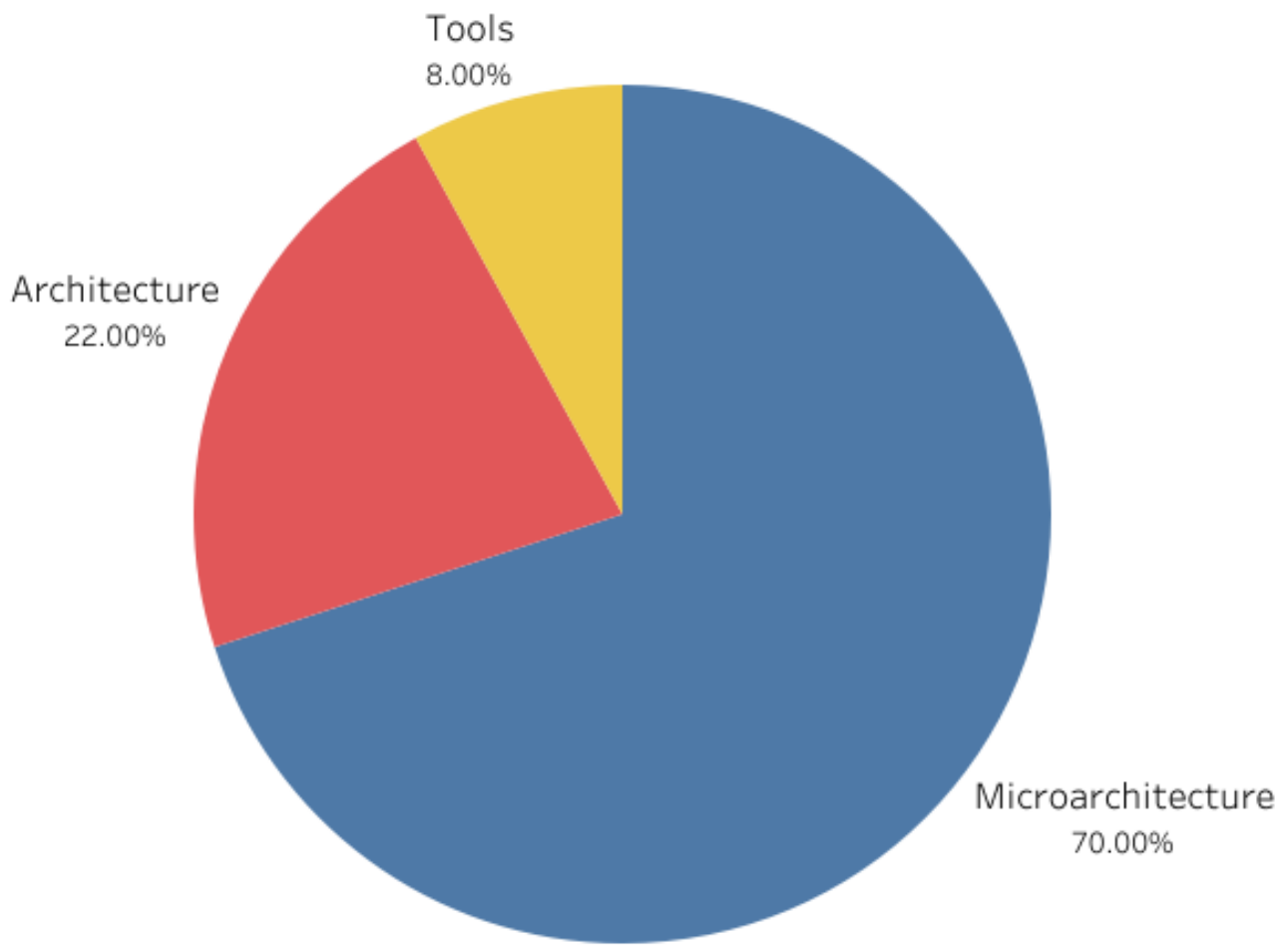

**Figure 9: Top 50 most cited ISCA papers of all time broken down by type.**

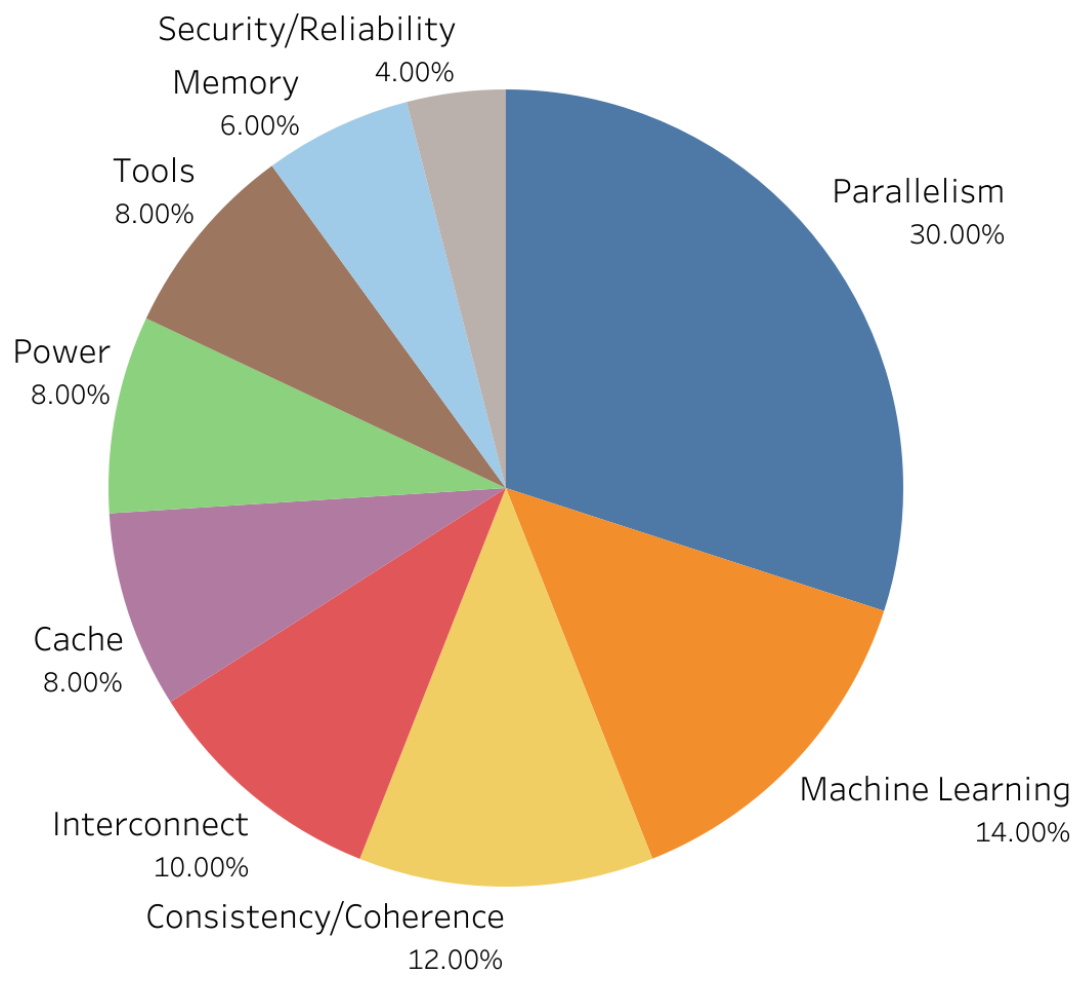

**Figure 10: Top 50 most cited ISCA papers of all time broken down by topic.**

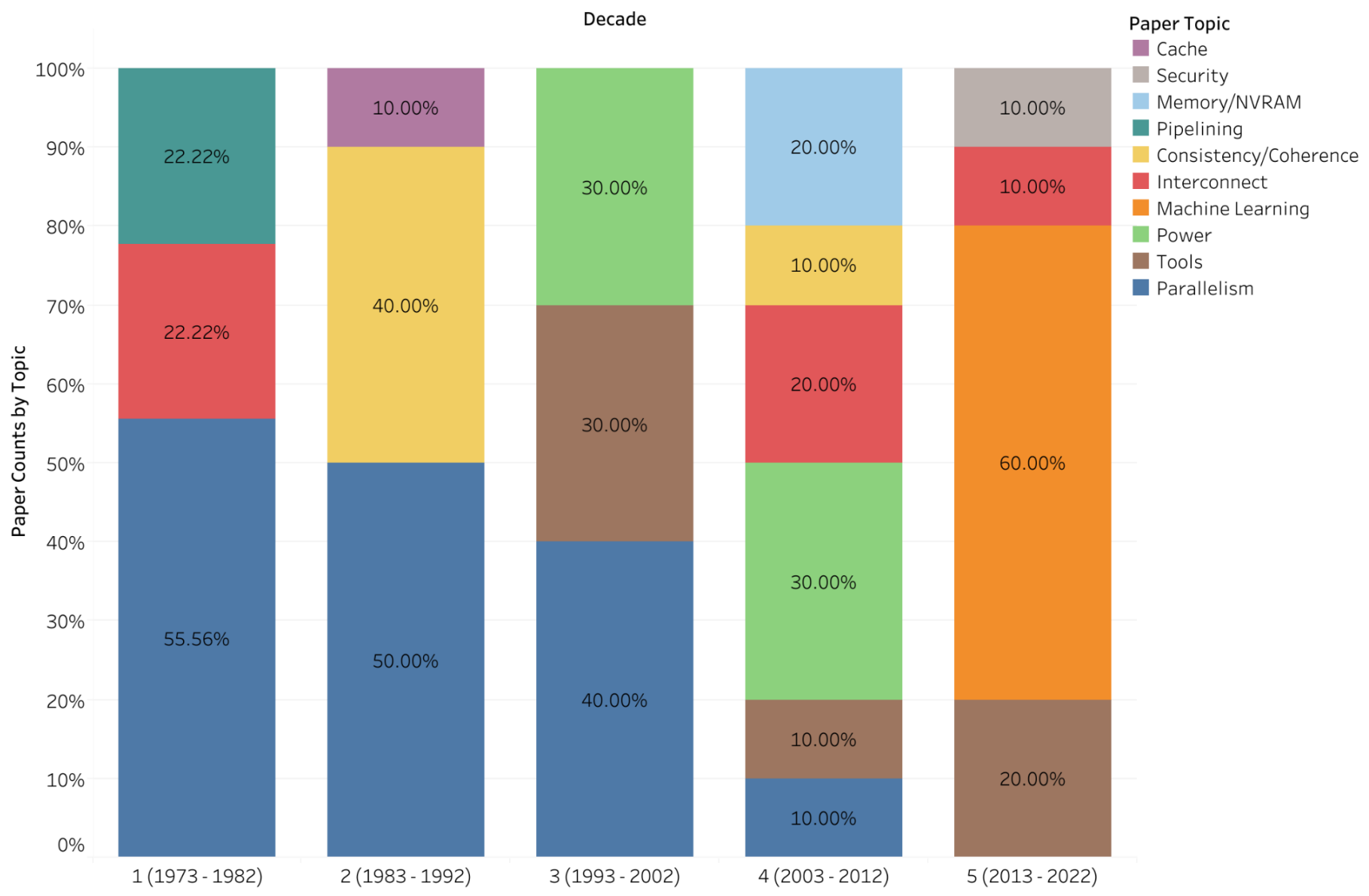

**Figure 11A: Topic distribution in each decade for per-year most cited ISCA papers**

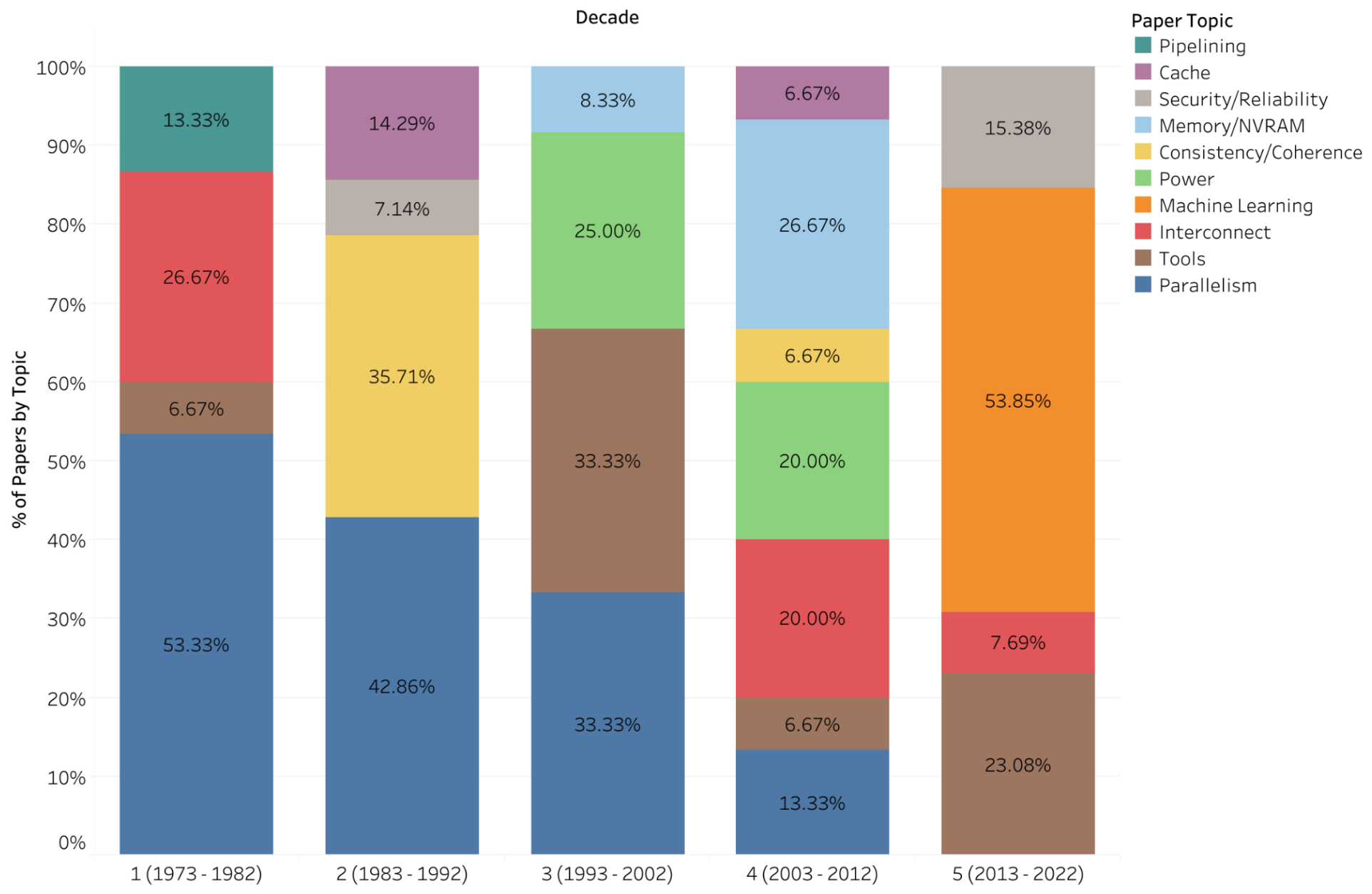

**Figure 11B: Topic distribution in each decade for papers with citation counts within 15% of top-cited paper each year.**

### 3.2.3 Topic Distribution over Decades

To further analyze how highly cited ISCA papers evolved over five decades, we show the % of top-cited papers distributed by topic for each decade in Figure 11. Figure 11A shows topic variation for the top-cited paper each year (Table 1), while Figure 11B includes the top-N papers per-year with citation counts within 15% of the top-cited paper that year. The trends in Figure 11A echo the detailed decade-wise analysis in Section 3.2.2. Figure 11B, which examines the trends at a finer granularity by including papers that were highly cited but were just shy of becoming the most-cited paper in a given year, also exhibits similar high level trends. *Parallelism* dominated the initial few decades, while in the third decade the number of papers with a focus on *Power* started to increase, and continued into decade 4. This also increased papers on simulators and benchmarks for power measurement, reflected in an increase in the percentage of *Tools* papers in the last few decades. Finally, topics such as accelerators and security appear much more prominently in the last decade, a trend also seen in the word clouds in Section 3.2.1.

### 3.2.4 Overall Top-Cited Papers

Examining the top cited paper each year provides broad trends about the most highly cited papers from each ISCA. However, it does not necessarily provide insight into the breakdown of those papers. Moreover, in some years there are multiple highly cited papers, which Table 1 cannot capture. Thus, Figures 9 and 10, as well as Table 2, analyze the Top 50 most cited ISCA papers by type and topic [10]. Our three types are:

1. *Microarchitecture*: Architecture techniques that could be used inside many computers;
2. *Architecture*: A description or proposal of a full computer architecture; and
3. *Tools*: Tools to help architects design computers, such as simulators or benchmarks.

Looking at the type mix in the pie chart in Figure 9, it's unsurprising that microarchitecture and architecture dominate the type of papers published at ISCA – after all these are and/or have been primary focuses of the Computer Architecture community in the past 50 years. However, it is interesting that tools that help architects makeup a non-trivial amount of the papers in the Top 50. In fact, the Top 50 list has only 3 tool papers, but 2 are in the top 5. These tools papers often have significant staying power. For example, the SPLASH-2, the highest cited paper thus far at ISCA, appeared 12 years before the next oldest paper in the top 10 cited papers. However, as mentioned in Section 2.1, this could also highlight the changes in citations and papers per year that bias these metrics towards more recent papers. These hurdles also make it highly more impressive that one paper from the first ISCA in 1973 made the Top 50 list (#42). Its ISCA Hall of Fame author was Jack Lipovski, who was also the first general chair of ISCA.

The architecture topics in Figure 11 provide a finer-grained classification; it uses the same classification scheme as the Top 50 to create a heatmap of topics over the years. As the chip design mantra today is power-performance-area, it is no surprise that parallelism and power are large slices. Machine learning (ML) accelerators have an unexpectedly large slice of the Top 50 cited ISCA papers, since the current excitement about deep neural networks started only a decade ago – the 7 ML papers in the Top 50 were published at ISCA between 2014 and 2017. ML papers in general are among the most highly cited in all of science and engineering [4]. For example, the ResNet ML paper from 2016 has 166,000+ citations [5]. Thus, the huge popularity of ML today likely accelerates citations to ML accelerators. Besides ML accelerators, topics like parallelism, coherence/consistency, power, and interconnects appear in at least 10% of the Top 50 cited papers – reinforcing the importance of these topics as discussed in Sections 3.2.1 and 3.2.2. Moreover, the heatmap in Figure 12 further demonstrates how these ISCA topics evolved over time. Some topics like ML appear heavily starting in 2015, while others like parallelism appear repeatedly across the 50 years, echoing the transition from multiprocessors to multi-core CPUs.

When comparing the per-year pie charts (Figures 7 and 8) to those for the Top 50 overall (Figures 9 and 10) highlights several interesting trends. Certain topics like parallelism (30%) and interconnects (10%) have the exact same proportion. However, microarchitecture (57%) and tools (12%) papers constitute a significantly different proportion of the per-year top cited papers than the overall Top 50 (70% and 8%, respectively). This potentially reinforces that papers, especially those from earlier versions of ISCA, may not be cited as much relative to newer ISCA papers, but still made significant impacts at the time. Moreover, it also shows how papers from recent years (e.g., on security) which have not yet had enough time to be cited enough to reach the Top 50, are also making significant impact.

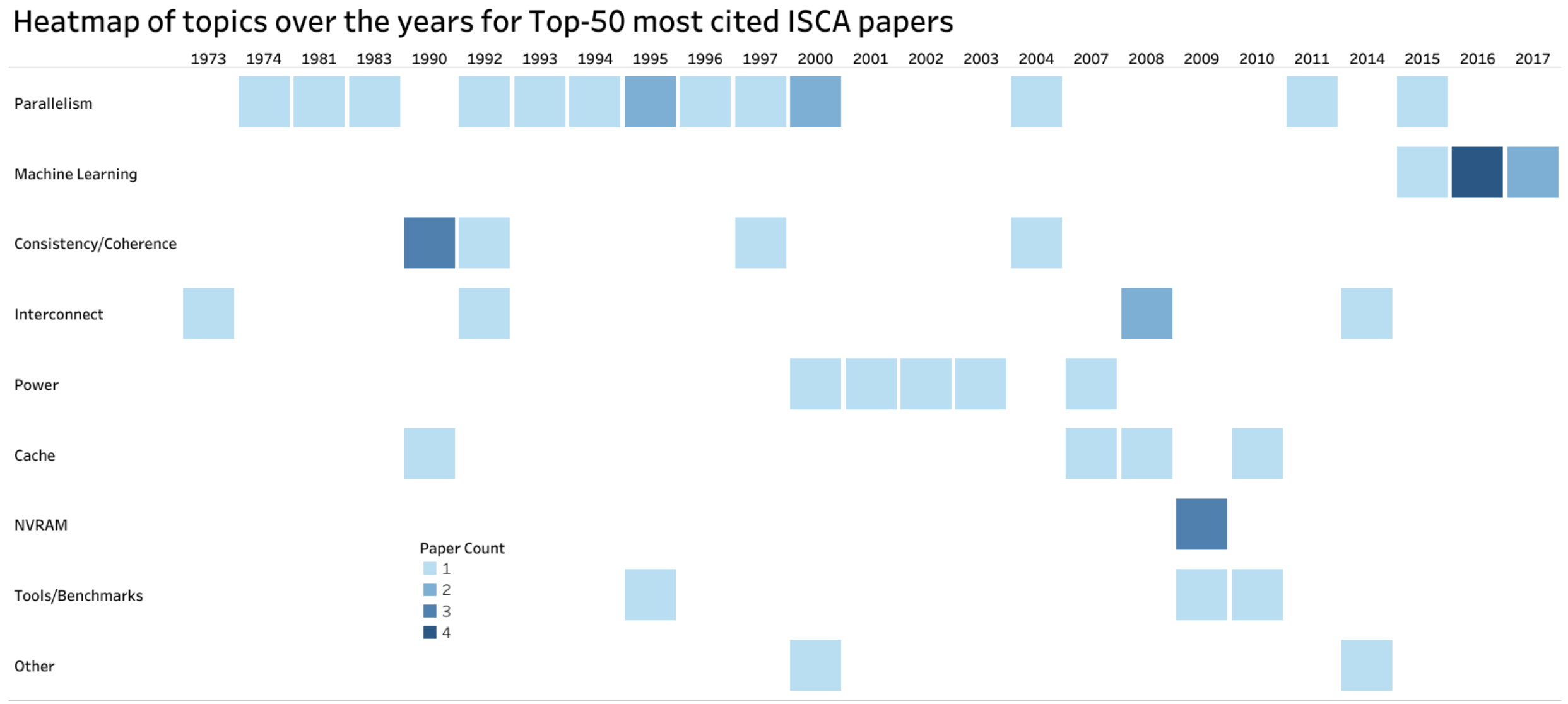

**Figure 12: Heatmap of how topics appeared over the past 50 years for the Top 50 cited ISCA papers.**

Finally, when examining specific authors who published these Top 50 papers, several authors stand out[2], including Norm Jouppi (Google), who won 2 Influential ISCA Paper Awards and co-authored 2 of the Top 15 and 5 of the Top 50. Others who co-authored several Top 50 papers are Joel Emer (5), Bill Dally (5), Anoop Gupta (4), Doug Burger (4), Kourosh Gharachorloo (3), John Hennessy (3), Onur Mutlu (3), and Dean Tullsen (3).

**Table 2: 50 Most Highly Cited ISCA Papers from first 50 Years of ISCA, including the year they were published, number of citations (as of May 2023), type, and topic.**

| Title | Year | # Citations | Type | Topic |
|---|---|---|---|---|
| The SPLASH-2 Programs: Characterization and Methodological Considerations | 1995 | 5361 | Tools | Benchmark |
| In-Datacenter Performance Analysis of a Tensor Processing Unit | 2017 | 4307 | Arch | Machine Learning |
| Wattch: a framework for architectural-level power analysis and optimizations | 2000 | 3840 | Tools | Power |
| Transactional Memory: Architectural Support for Lock-Free Data Structures | 1993 | 3390 | Micro | Parallelism |
| EIE: Efficient Inference Engine on Compressed Deep Neural Network | 2016 | 2727 | Arch | Machine Learning |
| Power provisioning for a warehouse-sized computer | 2007 | 2625 | Micro | Power |
| Active Messages: A Mechanism for Integrated Communication and Computation | 1992 | 2489 | Micro | Parallelism |
| Dark silicon and the end of multicore scaling | 2011 | 2405 | Micro | Parallelism |
| Simultaneous Multithreading: Maximizing On-Chip Parallelism | 1995 | 2353 | Micro | Parallelism |
| Improving Direct-Mapped Cache Performance by the Addition of a Small Fully-Associative Cache and Prefetch Buffers | 1990 | 2247 | Micro | Cache |
| Architecting phase change memory as a scalable DRAM alternative | 2009 | 1806 | Micro | NVRAM |
| Memory Consistency and Event Ordering in Scalable Shared-Memory Multiprocessors | 1990 | 1793 | Micro | Consistency/ Coherence |
| Scalable high performance main memory system using phase-change memory technology | 2009 | 1773 | Micro | NVRAM |
| ISAAC: A Convolutional Neural Network Accelerator with In-Situ Analog Arithmetic in Crossbars | 2016 | 1672 | Arch | Machine Learning |
| Temperature-Aware Microarchitecture | 2003 | 1644 | Micro | Power |
| Eyeriss: A Spatial Architecture for Energy-Efficient Dataflow for Convolutional Neural Networks | 2016 | 1578 | Micro | Machine Learning |
| PRIME: A Novel Processing-in-Memory Architecture for Neural Network Computation in ReRAM-Based Main Memory | 2016 | 1432 | Arch | Machine Learning |
| A reconfigurable fabric for accelerating large-scale datacenter services | 2014 | 1406 | Micro | Interconnect |
| The Turn Model for Adaptive Routing | 1992 | 1380 | Micro | Interconnect |

[2] We reiterate our comment from the introduction that impact can be measured on multiple dimensions, and this is only a partial view.

| | | | | |
|---|---|---|---|---|
| Multiscalar Processors | 1995 | 1354 | Micro | Parallelism |
| Memory access scheduling | 2000 | 1303 | Micro | Parallelism |
| Complexity-Effective Superscalar Processors | 1997 | 1285 | Micro | Parallelism |
| Drowsy Caches: Simple Techniques for Reducing Leakage Power | 2002 | 1221 | Micro | Power |
| Exploiting Choice: Instruction Fetch and Issue on an Implementable Simultaneous Multithreading Processor | 1996 | 1210 | Micro | Parallelism |
| A Study of Branch Prediction Strategies | 1981 | 1203 | Micro | Parallelism |
| The SGI Origin: A ccNUMA Highly Scalable Server | 1997 | 1202 | Arch | Consistency/ Coherence |
| Flipping bits in memory without accessing them: An experimental study of DRAM disturbance errors | 2014 | 1194 | Micro | Security |
| A durable and energy efficient main memory using phase change memory technology | 2009 | 1178 | Micro | NVRAM |
| Debunking the 100X GPU vs. CPU myth: an evaluation of throughput computing on CPU and GPU | 2010 | 1167 | Tools | Simulator |
| SCNN: An Accelerator for Compressed-sparse Convolutional Neural Networks | 2017 | 1122 | Arch | Machine Learning |
| ShiDianNao: shifting vision processing closer to the sensor | 2015 | 1081 | Arch | Machine Learning |
| The Stanford FLASH Multiprocessor | 1994 | 1052 | Arch | Parallelism |
| Transactional Memory Coherence and Consistency | 2004 | 1027 | Micro | Consistency/ Coherence |
| Lazy Release Consistency for Software Distributed Shared Memory | 1992 | 1022 | Micro | Consistency/ Coherence |
| Cache decay: exploiting generational behavior to reduce cache leakage power | 2001 | 1006 | Micro | Power |
| The Directory-Based Cache Coherence Protocol for the DASH Multiprocessor | 1990 | 994 | Micro | Consistency/ Coherence |
| Clock rate versus IPC: the end of the road for conventional microarchitectures | 2000 | 992 | Micro | Parallelism |
| Weak Ordering - A New Definition | 1990 | 980 | Micro | Consistency/ Coherence |
| Technology-Driven, Highly-Scalable Dragonfly Topology | 2008 | 967 | Micro | Interconnect |
| Adaptive insertion policies for high performance caching | 2007 | 940 | Micro | Cache |
| Banyan Networks for Partitioning Multiprocessor Systems | 1973 | 937 | Micro | Interconnect |
| 3D-Stacked Memory Architectures for Multi-core Processors | 2008 | 911 | Micro | Cache |
| High performance cache replacement using re-reference interval prediction (RRIP) | 2010 | 901 | Micro | Cache |
| A scalable processing-in-memory accelerator for parallel graph | 2015 | 882 | Arch | Parallelism |
| processing | | | | |
| Transient fault detection via simultaneous multithreading | 2000 | 873 | Micro | Reliability |
| Corona: System Implications of Emerging Nanophotonic Technology | 2008 | 869 | Micro | Interconnect |
| An analytical model for a GPU architecture with memory-level and thread-level parallelism awareness | 2009 | 863 | Arch | Tool |
| Single-ISA Heterogeneous Multi-Core Architectures for Multithreaded Workload Performance | 2004 | 860 | Micro | Parallelism |
| A Preliminary Architecture for a Basic Data Flow Processor | 1974 | 854 | Arch | Parallelism |
| Very Long Instruction Word Architectures and the ELI-512 | 1983 | 854 | Arch | Parallelism |

### 3.2.5 The Life of a Highly Cited Paper

Figure 13 shows the trends in per-year citations for the top-cited papers per-year in Section 3.2.2. We chose to examine these papers instead of the overall Top 50 papers, to address the bias discussed earlier about later-year papers getting more citations (and also because this made it simpler graphically!). The y-axis shows the number of citations for each year after publication. The dot closest to the x axis represents the first year after publication, the dot after that the year after, and so on. The size of the circles is proportional to the number of citations that year. Moreover, for a few of the early ISCA papers, we do not have citation history, so the top cited papers from those years are absent.

Although the citation paths are somewhat different for each of these papers, several common trends exist. We discuss those, as well as highlight a few example papers that stand out. First, many of the top papers from the 1970s and 1980s see slow, steady citations, often continuing 35+ years beyond their publication date. However, with the exception of Jim Smith's branch prediction paper in ISCA 1981 [26], these papers largely do not see the same large bursts of citations that papers from more recent ISCAs see. In comparison, top papers from ISCAs after 2005, including the ISCA 2009 Phase Change Memory paper [23], the ISCA 2011 Dark Silicon paper [18], Google's ISCA 2017 TPU paper [24], and others often see very large bursts of citations very early. Some of these papers continue to be cited highly today. As a result, these papers have reached very high citation counts and are some of the highest cited ISCA papers ever (Section 3.2.3). Collectively, these trends seem to further highlight the changes in citation policies and the increase in Computer Architecture papers per year (Section 2.1). Interestingly, in between these time frames, there are several top cited papers, including SPLASH-2 (the most cited ISCA paper ever, Section 3.2.3) that exhibit even more interesting trends. These papers see initial bursts of high citations, followed by a period of fewer citations, and are then cited highly again later on. We believe this highlights the prescience of some of these papers – as Dennard's Scaling ended and Moore's Law slowed, topics like parallelism (SPLASH-2 [32], Transactional Memory [33], ISCA 1993) and power consumption (Wattch [34], ISCA 2000) became even more popular, and thus these papers became highly relevant again later in life.

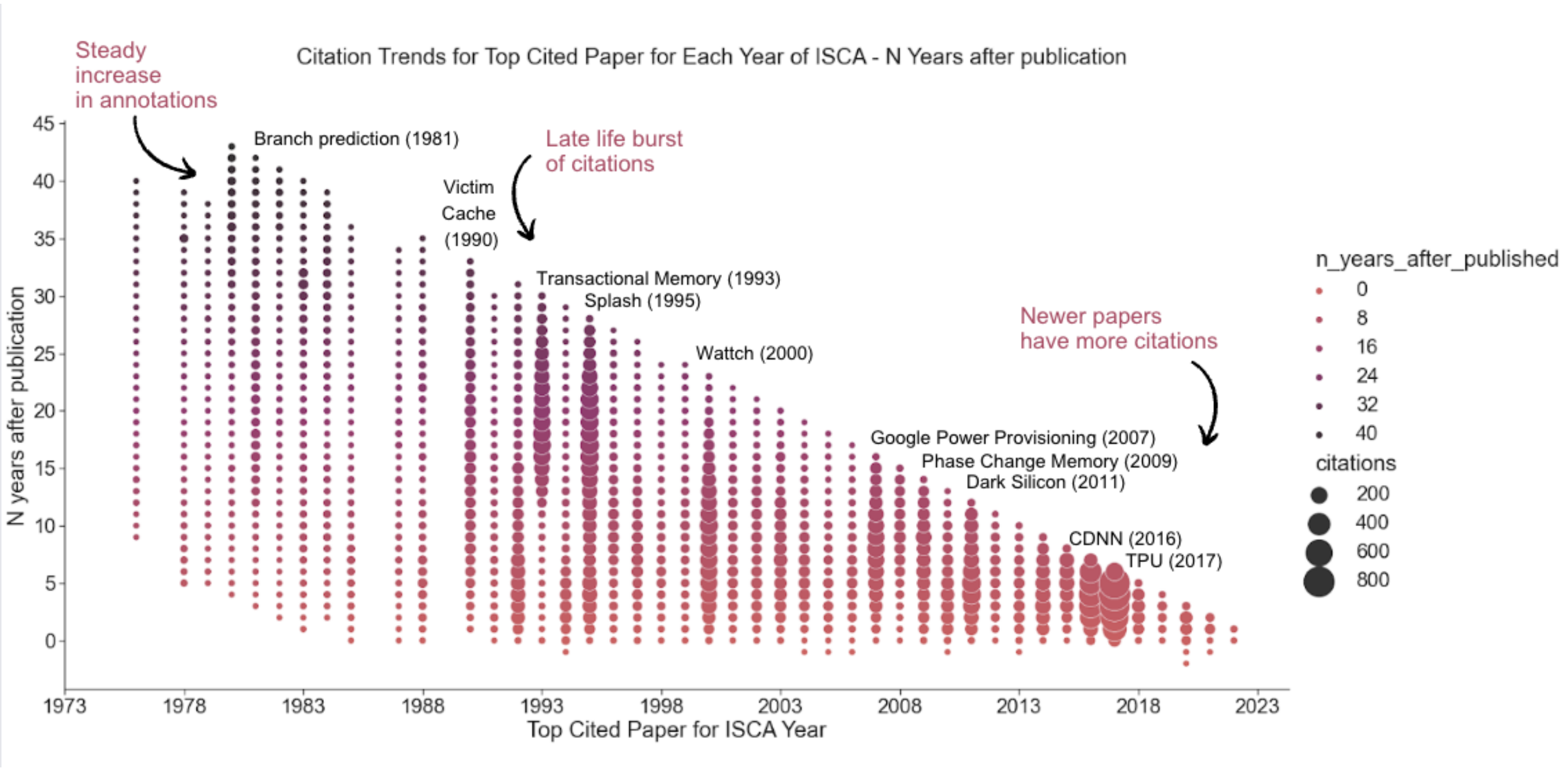

**Figure 13: Citation trends for top-cited paper of each year of ISCA. The y-axis shows the number of citations for each year after publication (proportional to the size of the circles). The gaps are because we do not have data for citation history for some of the earlier papers.**

## 3.3 Who is Publishing at ISCA?

### 3.3.1 Prolific Publishers at ISCA

In an effort to recognize top researchers submitting and publishing their work at ISCA, in 1995 Mark Hill and Guri Sohi created the ISCA Hall of Fame (HOF) [7]. Since 2020 this list has been updated and maintained by Matt Sinclair. Researchers are recognized as joining the ISCA HOF when they publish eight or more papers at ISCA. In the first 50 years of ISCA, 125 ISCA authors have reached this threshold. Within this group of distinguished architects, 82 of the 125 authors (66%) have published at least 10 ISCA papers and 15 of the 125 authors (12%) have published at least 20 ISCA papers. Figure 13 shows the histogram distribution per career length. The average HOF researcher has been publishing at ISCA for about 20 years, though we have a wide range of career lengths from less than ten years to more than 40 years. Arvind, Trevor Mudge, Dave Patterson, and Jim Smith have the distinction of having published at ISCA in all five decades!

Examining the diversity in the Hall of Fame also highlights the need to continue working on improving diversity in our community: while we don't have self-reported data, we used Gender API to obtain an estimate [17]. Based on this, we found that only 9% of HOF authors are female – confirming trends in prior work that the field skews heavily male [16]. Figure 14 further confirms this. It plots the number of female authors each year as a percentage of all authors, with the standard deviation representing how confident Gender API was in its classification. The percentage of female authors has increased over the years: from 3% (1973) to 15% (2022). However, this fraction is still much lower compared to male authors, suggesting that we need to continue our outreach efforts in this area.

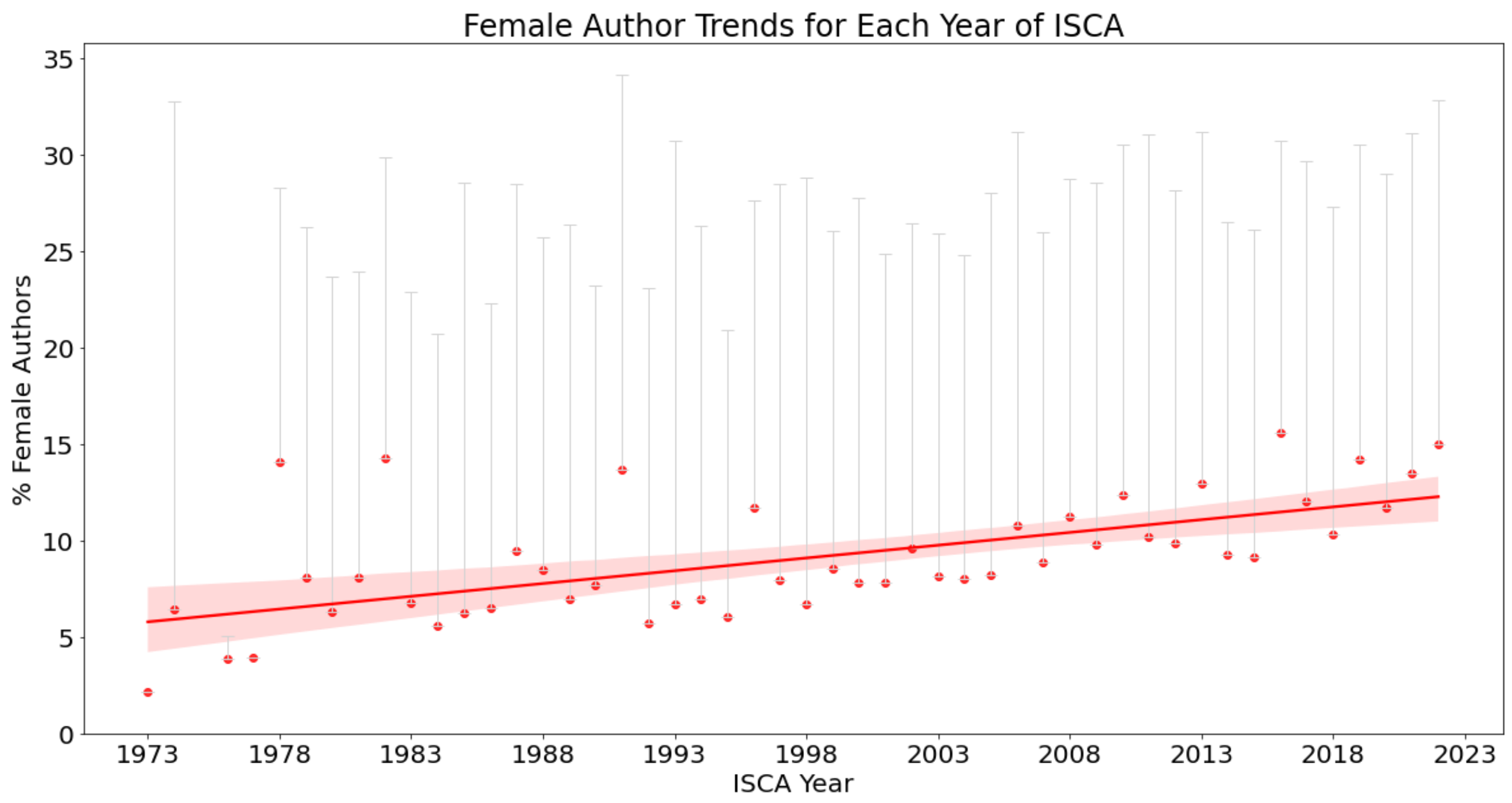

**Figure 14: Author trends that are identified as female per Gender API for every year of ISCA**

Moreover, of the 2134 papers published at ISCA in between 1973 and 2023, 1075 of them (50%) include one or more of the researchers in the Hall of Fame! This is remarkable, given that these 125 researchers only represent 2.5% of all ISCA authors in the first 50 years. However, prior work has shown that many papers being published by top researchers in the field is somewhat common across fields [14][15]. When further examining the breakdown of these authors between industry and academia, we find that about 14% of them work primarily in industry, 5% have mixed industry/academia affiliations, and 81% work primarily in academia. This authorship breakdown confirms and reinforces trends from a prior 2005 study by Peter Kogge [8]. One key difference though is that while early years of ISCA had more publications from industry, there was a time period where publishing industry papers (e.g., on "real" products) was challenging because of conflicting requirements between designing and selling processors versus expectations for disclosing information in ISCA papers. Hopefully, recent efforts to add an industry-specific ISCA track will help ensure that papers from both communities continue to be welcome in future ISCAs [11]. For example, the top cited paper from ISCA 2021 (the TPUv4i paper) appeared in this track.

### 3.3.2 Per-Decade Prolific Author Trends

We also further investigated the trends of how top publishers varied across the first five decades of ISCA. Below we list the "prolific authors" *per-decade.* In this context, "prolific" is defined as either having more than 5 papers in a decade (i.e., averaging a paper every other year at ISCA):[3]

**Decade 1 (1973-1982)**: Jack Lipovski, Howard Siegel, David Patterson, V. Carl Hamacher

**Decade 2 (1983-1992)**: Anoop Gupta, John Hennessy, Guri Sohi, James Goodman, Janak Patel, Mark Hill, Yale Patt, Andrew Pleszkun, Mark Horowitz, David Patterson, Edward S. Davidson, H. T. Kung, Susan Eggers, Wen-Mei W Hwu, Anand Agarwal, Henry M Levy, Jean-Loup Baer, Mary K Vernon, Randy H Katz, William J Dally

[3] Here we order "prolific" authors first by number of publications in that decade, and second by alphabetical order.

**Decade 3 (1993-2002)**: David Wood, Guri Sohi, Andre Seznec, Norm Jouppi, Steve Reinhardt, Yale Patt, Babak Falsafi, Brad Calder, Dean Tullsen, Dirk Grunwald, Wen-Mei Hwu, Doug Burger, Jaswinder Pal Singh, John Paul Shen, Josep Torrellas, Kai Li, Kourosh Gharachorloo, Mark D. Hill, Trevor N. Mudge, Anant Agarwal, Anoop Gupta, James E. Smith, Joel S. Emer, Luiz André Barroso, Margaret Martonosi, Mark Horowitz, Michel Dubois, Per Stenström, Sarita V. Adve, Shubhendu S. Mukherjee, Susan J. Eggers, Todd M. Austin, William J. Dally

**Decade 4 (2003-2012)**: Onur Mutlu, Josep Torrellas, T.N. Vijaykumar, Christoforos Kozyrakis, Babak Falsafi, Yale N. Patt, Mark D. Hill, Mikko H. Lipasti, Norman P. Jouppi, Thomas F. Wenisch, William J. Dally, David A. Wood, Doug Burger, Frederic T. Chong, John Kim, Krste Asanovic, Luis Ceze, Moinuddin K. Qureshi, Naveen Muralimanohar, Sanjay J. Patel, Timothy Sherwood, Trevor N. Mudge, Amir Roth, Anastasia Ailamaki, Chita R. Das, James E. Smith, Joel S. Emer, Karin Strauss, Mark Oskin, Milo M. K. Martin, Narayanan Vijaykrishnan, Parthasarathy Ranganathan, Rajeev Balasubramonian, Yuan Xie, Al Davis, Anand Sivasubramaniam, Benjamin C. Lee, Changkyu Kim, Christopher J. Hughes, Dennis Abts, Engin Ipek, Gabriel H. Loh, Hyesoon Kim, Karthikeyan Sankaralingam, Lizy Kurian John, Mark Horowitz, Mateo Valero, Mattan Erez, Michael C. Huang, Pradip Bose, Scott A. Mahlke, Stephen W. Keckler, Steven K. Reinhardt, Sudhanva Gurumurthi, Todd M. Austin, Zeshan Chishti

**Decade 5 (2013-2022)**: Onur Mutlu, Josep Torrellas, Yuan Xie, Nam Sung Kim, Jeremie S. Kim, Christoforos E. Kozyrakis, Frederic T. Chong, Hadi Esmaeilzadeh, Jangwoo Kim, Jason Mars, Lingjia Tang, Sreenivas Subramoney, Xuehai Qian, Alberto Ros, Christopher W. Fletcher, Gennady Pekhimenko, Jae W. Lee, Mahmut T. Kandemir, Margaret Martonosi, Minesh Patel, Moinuddin K. Qureshi, Nandita Vijaykumar, Satish Narayanasamy, Stefanos Kaxiras, Won Woo Ro, Yan Solihin, Abhishek Bhattacharjee, Carole-Jean Wu, Chita R. Das, Daniel Sánchez, David Brooks, David T. Blaauw, Jayesh Gaur, John Kim, Kunle Olukotun, Mattan Erez, Murali Annavaram, Rakesh Kumar, Reetuparna Das, Saugata Ghose, Thomas F. Wenisch, Tor M. Aamodt, Aamer Jaleel, Abdullah Giray Yaglikçi, Amir Yazdanbakhsh, Anand Sivasubramaniam, Antonio González, Ashish Venkat, Daniel A. Jiménez, Dean M. Tullsen, Eiman Ebrahimi, Gabriel H. Loh, Gu-Yeon Wei, Mengjia Yan, Minsoo Rhu, Ronald G. Dreslinski, Timothy Sherwood, Tony Nowatzki, Tushar Krishna, Amro Awad, Babak Falsafi, Bhargava Gopireddy, Daniel Lustig, David W. Nellans, David Wentzlaff, Hasan Hassan, Jaehyuk Huh, Jian Huang, Joel S. Emer, Juan Gómez-Luna, Jung Ho Ahn, Krste Asanovic, Leibo Liu, Lieven Eeckhout, Lois Orosa, Mike O'Connor, Myoungsoo Jung, Natalie D. Enright Jerger, Phillip B. Gibbons, Shaojun Wei, Shouyi Yin, Simha Sethumadhavan, Stephen W. Keckler, Sudhakar Yalamanchili, Tae Jun Ham, Tao Li, Tianshi Chen, Yaqi Zhang, Yoav Etsion, Yuhao Zhu

As before, there are several caveats to these lists including the large number of papers in later years as well as artifacts from using ISCA-decades as a cut-off epoch. However, the data is still interesting in reflecting both a growing vibrant community as well as an increasing diversity in authors including new authors every decade.

### 3.3.3 The Path to the ISCA Hall of Fame

Figures 15 and 16 show additional details about the path to the ISCA HOF for the 125 current members. As discussed in Section 3.3.1, Figure 15's histogram shows the average HOF member has a "to-date ISCA career[4]" (the delta between their first paper and their most recent paper) of 20 years. However, the range is significant: as few as 8 years and as many as 46 years. This is highlighted by Figure 16, which shows how an ISCA HOFers number of papers (on the y-axis) increases as their career advances (ISCA number on the x-axis). The x-axis maxes out at 45 which, as we discussed, is the longest tenure for any hall-of-famer currently.

[4] The term "career" might be a misnomer here since the hall of fame includes authors who are currently active with a potentially long publication record at ISCA still ahead.

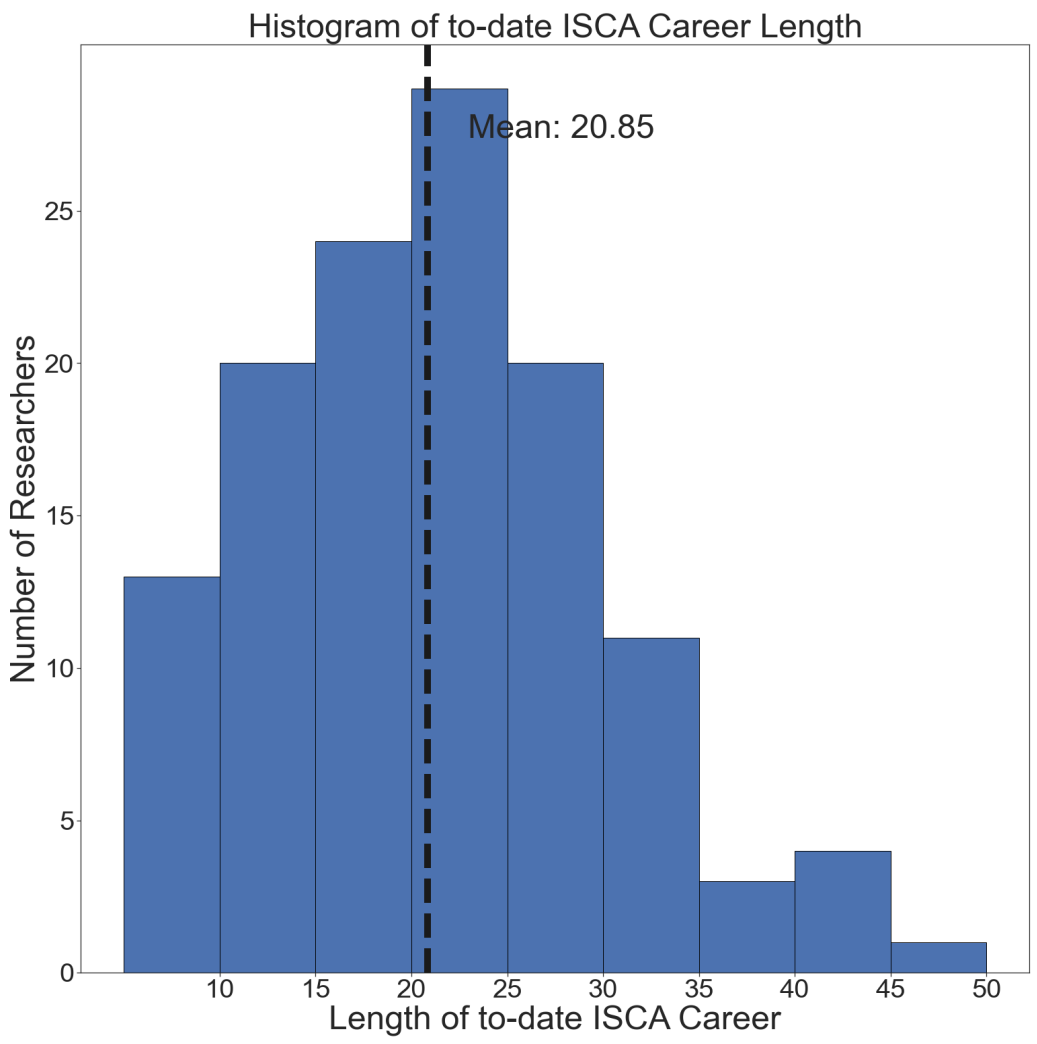

**Figure 15: Histogram of average length of ISCA career for the members in the ISCA HOF. The lowest tenure for the HOFers was 8 years, and the median was ~21 years. The maximum was 46 years!**

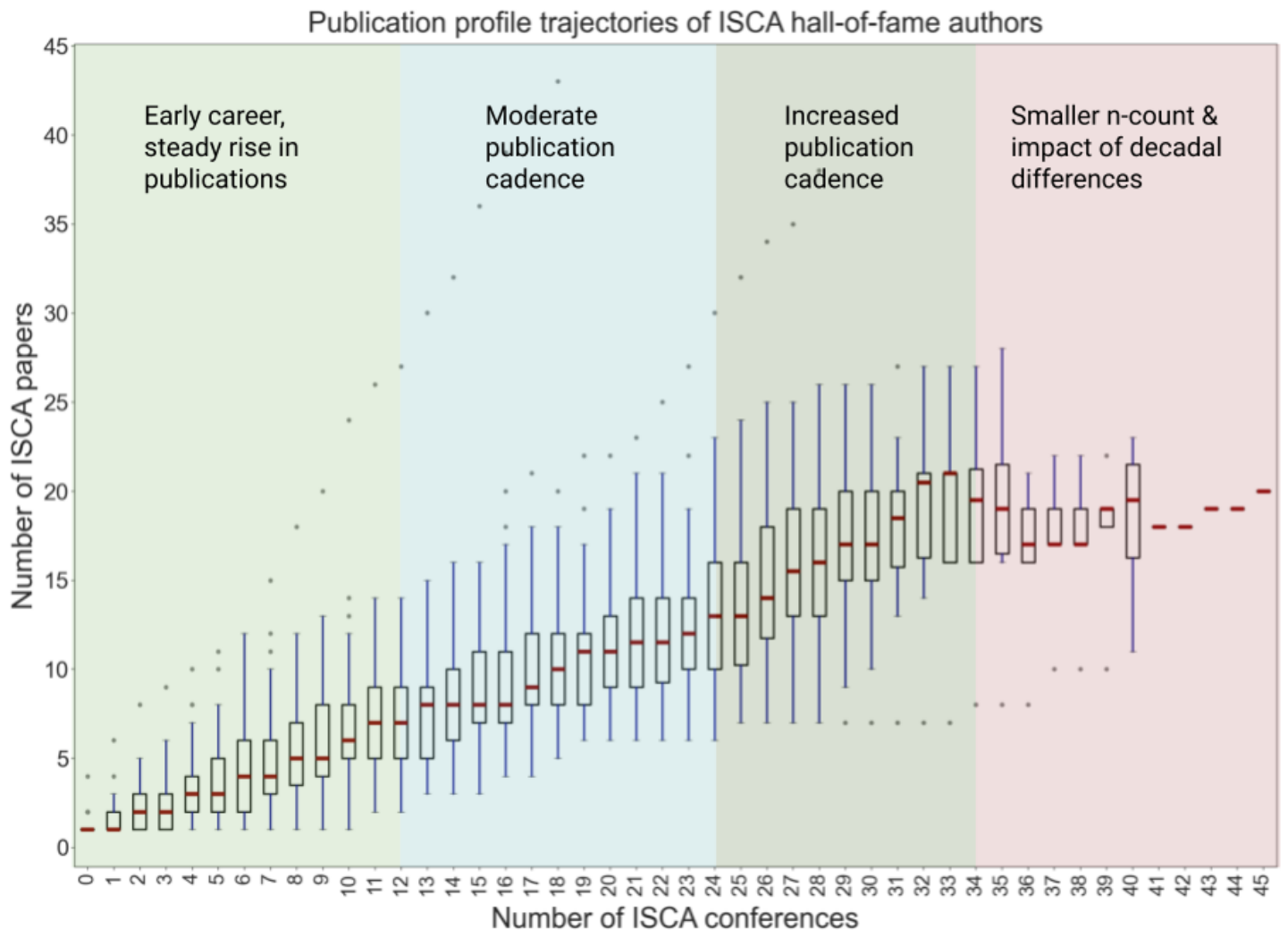

**Figure 16: The average career of an ISCA hall-of-fame author, represented as the growth in number of ISCA papers over the ISCA career. The "o" signs represent outliers, the red lines represent the medians, and the lower and upper ends of the boxes represent the 25th and 75th percentiles respectively.**

For most HOFers, we see that there are three distinct phases: a quick rise in ISCA paper count in their early career, a period of slow, but steady growth in ISCA paper count after reaching the HOF, and the finally later in their careers many HOFers see another increase in ISCA papers. We suspect that this last phase represents additional papers occurring through collaborations or students. Nevertheless, it is important to note that the variation is large here – especially in later years. We believe this also represents the trend where the time to reach the HOF has been decreasing in recent years (similar to how more authors appear in the later decades of Section 3.3.2).

## 3.4 ISCA Memories

To add a qualitative dimension to our retrospective, we also surveyed the community to get their memories of ISCAs over the decades. We emailed all 123 living members of the ISCA Hall of Fame and asked each of them three questions: (1) their favorite ISCA-related memory, (2) for any highlights from the decade when they first attended ISCA, and (3) for their favorite ISCA paper of all time. While many wisely declined to answer the third question, we collected a treasure trove of memories from attending the first 50 ISCAs.

The overwhelming impression that these stories yielded is that, in addition to the technical component, the respondents really valued the community and social aspect of ISCA. The most storied years were destinations in Europe and Asia: we heard many times about the banquet and its afterparty in Barcelona in 1998, the exquisite setting of ISCA 2010 in the French port town of Saint-Malo, and, more recently, the food and entertainment in Seoul in 2016. ISCA researchers remembered when they met each other for the first time, when the banquet featured particularly good food, and when the conference gifts were especially useful. One attendee of ISCA 1996 in San Diego, for example, reports that she still uses the flip-flops from that year, which leave "ISCA-34" footprints on the sand behind you when you walk on the beach. In aggregate, the stories depict a long history of adding an in-person, interpersonal dimension to the research that makes up our field. And while it was only one, we did even hear from one ISCA attendee who has positive memories from a "virtual ISCA" in 2020 during the height of the COVID-19 pandemic.

Another common theme recounted researchers' first-ever talk at ISCA. Some of these memories were purely fond ones, but more often they were memories of the nervous moments before the talk. That intimidating moment when you see famous and respected senior researchers in the audience seems to be a constant across the decades—even for ISCA denizens who now, decades later, play that same role for new authors at ISCA in 2023. The set of stories we collected contained at least one "I-first-met chain": when researcher *A* fondly remembered meeting their research hero, researcher *B*, at their first ISCA while, separately, researcher *B* submitted a nearly-identical story about their excitement upon meeting researcher *C* at their first. The stories highlight a key function of ISCA that has remained constant through its five decades: connecting new entrants to the field with established researchers to coalesce a community. For more specific examples, we refer the reader to the special series of social media posts we launched as part of the #ISCA50 celebratory campaign.

# 4. Conclusion

Overall, our study highlights the dynamic nature of ISCA over the first 50 years. Topics have changed, the community has grown, but a constant across five decades has been the growing impact and diversity[5] of ISCA. We find that as computer systems have evolved, ISCA's topics have evolved with it: including minicomputers, general-purpose uniprocessor CPUs, multiprocessor and multi-core CPUs, general-purpose GPUs, and accelerators. Given how these innovations have helped enable significant societal changes, we believe this highlights the importance of Computer Architecture and the need for continued innovation in the next 50 years of ISCA.

Looking ahead, the future looks exciting. Based on current trends, it seems likely that topics like accelerators, biological-inspired computing, carbon-aware computing/environmental sustainability, edge computing, ethical computing, quantum computing, and security/privacy will play important roles in the next couple decades. Moreover, questions abound about the future of transistors in the next few decades as Moore's Law continues to fade [9], it seems likely that subsequent decades of ISCA will need to rethink how systems should be designed with potentially very different underlying computational substrates. This also highlights the need to continue innovating around new technologies and examining how to effectively break abstraction layers (including compilers, operating systems, VLSI, and others) in order to design efficient future architectures. One notable interesting use-case is the application of AI/ML in the *design* of computer architectures.

Regardless of the changes ahead, researchers leading the charge on these future designs must learn from the past in order to best identify how future systems should be designed. Here, it seems a proper tribute to leave the last words to Jack Lipovski, whose observations from 50 years ago remain just as relevant today:

> *This symposium may well be, in the hind-sight of ten years from now, a marked turning point in Computer Architecture. With the dissolution of the Spring and Fall Joint Computer Conferences[6], one of the major forums for Computer Architecture has been lost. So we have begun an annual symposium on Computer Architecture, to be rotated from year to year around the world. …*
>
> *The papers in the symposium indicate the growth of Computer Architecture as a science. Although it is difficult to explain the reasoning behind the decisions made in an architecture, in particular, the architecture of a practical machine, this reasoning is the basis of a science. …*
>
> *We hope that attendees will emphasize questions on the reasoning behind the architecture, and the authors will prepare for such questions. If this becomes a tradition in this annual symposium, it should orient authors toward the scientific explanation of their architectures for later symposia.*

---

[5] Interestingly some of this diversity reflects even in the authorship of this document. To collect, analyze, and parse this data required a team effort of academic and industry researchers ranging from high schoolers, undergrads, grad students, pre-tenure faculty, post-tenure faculty, emeritus faculty, as well as junior industry and senior industry practitioners! Note: recognizing the equal value of diverse contributions, the author list is ordered reverse alphabetically by last name, sorted into student and non-student authors.

[6] Starting in 1951, the SJCC and the FJCC conferences covered all of computer science and reached 15,000 attendees at their peak [6]. The Federated Computing Research Conference (FCRC), founded by the Computing Research Association in 1993, is an attempt to recreate the benefits of large multidisciplinary meetings like the SJCC and the FJCC by colocating several single discipline conferences at the same venue and at the same time. ACM now holds FCRCs every 4 years.

*— Jack Liposvki, Preface of the First Annual Symposium on Computer Architecture Proceedings, University of Florida, December 9–11, 1973*

## Appendix A: ISCA Memories

The following is a summary of the memories collected from members of the ISCA community.

# Todd Austin

Favorite Memory:
“ It has to be my first conference presentation, which occurred at ISCA-19 (1992), where I was presenting my first paper with my advisor Prof. Guri Sohi. I was unbelievably nervous for the whole affair and completely unsure of myself, but I was taking some solace in that I didn't know anybody in my session. It was a rare early multitrack ISCA conference in Gold Coast Australia. About two minutes before my talk was to start, I heard clapping through the walls from the other session, and a few moments later... David Patterson, John Hennessy and Arvind all walk in together to attend my portion of the session. Oh god... Never. More. Nervous! All I could think of was that if I screw this talk up it will be the end of my career! The rest of the session was a bit of a blur. Fortunately, I memorized what I was going to say for my first couple of slides, and then I managed to complete the talk. After my talk, Prof. Gus Uht from University of Rhode Island approached me and told me how much he enjoyed my talk. I will never forget that moment!

Close second most memorable ISCA memory (I tell this story in every computer architecture class I teach :) ): It was ISCA 1997, and Robert Tomasulo was receiving the Eckert–Mauchly Award for his contributions to dynamic instruction scheduling. After the award ceremony, there was an informal receiving line to meet Robert, and I joined the line with Prof. Gary Tyson (of FSU). We waited for what seemed like one-half of an hour to get to the front of the line. While in line, I was fretting over what question would I ask Robert, since this would surely be the one and only time I'd ever get to talk to this giant of computer architecture. Perhaps I'll ask him how they modeled their design before building it, or I might ask him if they were thinking about the possibility of register renaming at that time... Finally, Gary and I got to the front of the receiving line and Robert looked at me and said "Sorry kid, I need a cigarette, I'll talk to you later." And that was my one brush with the giant of computer architecture named Robert Tomasulo!"

## Ramon Belvide

Favorite Memory:
"Together with my colleagues J.L. Balcazar, E. Herrada and J. Labarta, 36 years ago in 1987, we published at ISCA'14 the paper "Optimized Mesh-Connected Networks for SIMD and MIMD Architectures". It contained results from my thesis, which extended ideas presented by Carlo Sequin in his ISCA'8 paper "Doubly Twisted Torus Networks for VLSI Processor Arrays", published in 1981. Interestingly, the networks used by Google TPUs are three-dimensional evolutions of these seminal ideas. They are described in the paper "TPU v4: An Optically Reconfigurable Supercomputer for Machine Learning with Hardware Support for Embeddings", which will be presented next June in the Industry Session of this ISCA'50.

I have to confess that our ISCA'87 paper was on the verge of not being published, as Txiki, my water dog, destroyed the author's kit almost completely. Those were the old times when camera-ready versions were processed on huge sheets of paper. During my teaching hours at the UPC, I usually left Txiki alone in my office. Upon returning after class one day, imagine the shock to find shreds of paper all over. And imagine the horror, when the shreds were identified as the sheets of the final version of the camera ready ISCA paper that had to be sent by air mail from Barcelona. Only one of the eight pages was usable. Fortunately, with the support of my friend Victor Vinyals and the reprographic office of the School of Architecture of the UPC, the paper was restored and ultimately published without detectable problems. This funny anecdote is registered in the Proceedings of ISCA'14, as stated in the (dis)acknowledgements of our paper:

Acknowledgements:
We are grateful to Celestí Rosselló for his kind help with the preparation of the figures, and to prof. Mateo Valero and other colleagues in our Departments for their helpful suggestions and encouragement. On the contrary, the first author's dog 'Txixi' is to be blamed if the quality of the copy is bad, since he ate most of the author's kit."

## Jack Cook

Favorite Memory:
“I presented "There's Always a Bigger Fish" at ISCA in 2022 with Jules Drean, and it was my first research paper and conference presentation. It was an amazing experience for me, and I'm incredibly grateful to all the reviewers, attendees, and conference staff who made it all possible! Thanks ISCA!”

## Babak Falsafi

Favorite Memory:
**“**Steve Reinhart's Typhoon talk in ISCA'94 with FLASH, Typhoon and Alewife. Steve tried to break the ice with a table comparing the three systems at the end of the talk, with a last row indicating which machine had a T-shirt for its team members (at ISCA) with FLASH and Alewife students all wearing their project t-shirts and Typhoon not having any t-shirts. Right after this talk, the first person at the mic was John Hennessy asking the question about who needs user-level software coherence protocols and reminding people of proposals for microprogramming and how user-level microcode may have been exciting as an idea (which eventually perished) decades before.**”**

## Jayesh Gaur

Favorite Memory:
“My favorite ISCA memory was the amazing dinner party at Korea, ISCA 2016 with a fabulous hip-hop dance performance (Maybe, it was K-pop). The whole audience of computer architects () were just in awe looking at that performance. A lot of fun - also to mention the Korean food (for vegetarians like me), that was just too good to go with it. ”

## Mark D. Hill

Favorite Memory:
“ISCA 1998 was in Barcelona hosted by Mateo Valero and others. After a wonderful late outdoor banquet, some of us convinced the organizers to let many of us commandeer a conference bus to take at the waterfront for after party. (It takes a big time zone change for an early riser like me to ever attend an after party.) We had a good time there, including playing miniature golf on a rooftop of a bar/disco.”

## Daniel A. Jimenez

Favorite Memory:
“One of the things I remember about previous ISCAs is the gifts. The weirdest and one of my favorites was ISCA 2002 in Anchorage, Alaska. That happened to be my first ISCA, too. Yale Patt and Dirk Grunwald were general chairs. They gave everyone a knife, specifically an Ulu knife which is a traditional kitchen implement in native Alaskan culture, but it looks like a Klingon weapon. This was less than a year after 9/11, which was the very worst time to give out knives to people about to get on airplanes back home. People who had only brought carry-on luggage had to check their knives or leave them in Alaska. Still, it is a pretty cool knife and I still have it in my office.

Another really good ISCA in terms of gifts was ISCA 2010 in Saint-Malo, Brittany, France. André Seznec was general chair and he put on a great conference. There were a lot of gifts, but two that stand out in my memory are a black beach towel embroidered with the ISCA logo, and these extremely good caramels "au buerre salé" i.e. "with salted butter." These are literally the best sweets I have ever had in my life. Over a decade later, a friend who knew how much I liked them asked André to bring me some more of those caramels from Brittany when he came to ISCA 2022. I still have a few left.

You might get some tales of debauchery that may or may not be appropriate. The one from ISCA 2005 in Madison might be wholesome enough to provide some entertainment. Apparently it's a tradition for students at the University of Wisconsin to go to the Essen Haus and play a drinking game where a glass boot is filled with beer and passed around the table. You have to take a drink. If you are the next-to-the-last person to empty the boot, you have to pay for the next one, so you're motivated to drink as much as possible, especially if the boot is near empty. Sounds like harmless fun. Well, we did this at ISCA with a large number of people but didn't take into account that passing around a drinking vessel to a bunch of people who have come from all over the globe and brought with them a large variety of international microbes was just asking for trouble. Many of us got really sick over the next few days (and some of the less hardy ones were sick that night). I had to give a talk at PLDI the next week (which was in Chicago, so driving distance from Madison) and I was really sick when I gave that talk.

# Nam Sung Kim

Favorite Memory:
“With my first ISCA paper, Drowsy Caches in 2002, I attended ISCA for the first time, and this brought me my first ISCA Influential Paper Award in 2017 where all my family attended together to celebrate winning this award.”

# Jingwen Leng

Favorite Memory:
“This was my first paper, which is fortunately to be ISCA 2013 (ten years ago)!”

# Margaret Martonosi

Favorite Memory:
“My first ISCA paper was in 1996.

Informing memory operations: providing memory performance feedback in modern processors
Mark Horowitz, Margaret Martonosi, Todd C. Mowry, Michael D. Smith
ISCA '96: Proceedings of the 23rd annual international symposium on Computer architecture. May 1996
https://dl.acm.org/doi/10.1145/232973.233000

I was a young faculty member at Princeton. (My grad school publications were mostly in Sigmetrics.) Note that there were no grad students on this paper. We four professors all knew each other from our prior time Stanford, but worked on this remotely together from other schools. We each brought in different strengths: Mike Smith and Todd were strong on microarchitecture and compilers. I did memory hierarchies and performance monitoring.

Mark Horowitz was great at microarchitecture and circuits. The early version of this work was rejected from ISCA 1995, and we waited another full year to resubmit a revised version. (In contrast to today's world where papers often get resubmitted very very soon to the next deadline.)

25+ years later, I am still puzzled while we haven't allowed memory hierarchy behavior to be better exposed to software so we could act on it more intelligently.

—

About ISCA 1996 itself: That was part of the second FCRC, which was in Philadelphia. We from Princeton took a van down each day to be able to bring lots of students. That was a presidential election year in the US, and I recall that some of our colleagues met and shook hands with Republican Presidential candidate Sen Bob Dole in the Reading Terminal Market food court near the conference in the Philadelphia Convention Center across the street. Dole had heard about FCRC and was eager to meet the computer folks in town: https://en.wikipedia.org/wiki/Bob_Dole_1996_presidential_campaign

—

On a completely different note: I somehow still have and still use the ISCA-34 flipflops given out as swag when ISCA was in San Diego in 2007. They have ISCA-34 cut into the soles so that you leave ISCA-34 footprints on the beach when you walk. Thank you Dean Tullsen, ISCA-34 General Chair! For a while I only used them at the beach. I have now upgraded them into everyday use at the pool. Somewhere in my many photos, I have photos of them at the NJ beaches from ocean swimming races I have done. I haven't managed to find any of these photos but will keep looking. I am tempted to wear them to the 50th birthday ISCA celebration panel this year. If I find any of the beach photos, I'll upload that to share."

My clearest memory of the conference itself is Jim Thornton's invited talk. Thornton was a hero of mine. He had been the lead designer of the CDC 6600 central processor, the chief architect for the Star-100 (a pioneering vector machine), and the founder of a pioneering LAN company that eventually lost out to Ethernet. He walked to the front of the auditorium, no podium, and gave a well-organized speech without the aid of notes. Old-school.

Having a paper at ISCA and the encouragement I got from other attendees gave me the confidence to change research areas, so in the Fall of 1981, I left CDC and returned to the University of Wisconsin, primed for a career in high performance computer architecture.

For the 8th ISCA, 90+ papers were submitted and 40 were accepted, 13 of which had a single author. The program committee was composed of 22 men. My submitted paper received two reviews, one of which pointed to two related papers and the other was a single sentence stating that the paper was good, but not excellent. The publications chair worked at IBM in Rochester and thought it would be a good idea if all the papers were in a "uniform and professional style". So someone in the publications department at IBM re-typed all of the papers, thereby producing an especially thick proceedings. I suggest you get a hard copy of the proceedings, leaf through all the pages for the tactile pleasure, and check out the state of computer architecture 42 years ago."

# Joshua San Miguel

Favorite Memory:
"Two wonderful events that I was lucky to co-organize:

1) ISCA Hockey 2017 in Toronto at the historic Maple Leaf Gardens. Photo credit: Julie Hsiao.
2) The first in-person Undergrad Architecture Mentoring (uArch) Workshop after the pandemic at ISCA 2022; an amazing cohort of undergrads from all over the world."

## Yunho Oh

Favorite Memory:
"The ISCAs that I presented my papers"

## David Patterson

Favorite Memory:
"At ISCA 22 in 1995, going to dinner at a small restaurant in Cinque Terre Italy with several colleagues, eating a great meal and drinking Limoncello (a fruit liqueur) ."

## Yubin Qin

Favorite Memory:
"I've been very fortunate this year. My first first-author paper was accepted by ISCA. I remember four years ago when I first started my research journey, I was deeply inspired by an ISCA paper and had this amazing feeling of a time-space intersection."

## Parthasarathy Ranganathan

Favorite Memory:
"One of my favorite memories of ISCA was the time I met Sir. Maurice Wilkes. I was a graduate student at that time and was in awe of him, but when I and a fellow graduate student went to say hi to him, he was nice enough to share some very interesting comp.arch anecdotes with us and we even had coffee/drinks together!

Another of my memorable ISCA moments was when I gave my keynote at ISCA. I had my advisor (Sarita Adve), her advisor (Mark Hill), and his advisor (David Patterson) all sitting in the first row. Three generations of academic advisors! Talk about pre-talk jitters! (And my talk was titled "The end of Moore's law or a computer architect's mid-life crisis?")"

## Andre Seznec

Favorite Memory:
"I was GC of ISCA2010 in Saint-Malo. The ones that do not live in Brittany and only attended ISCA in Saint-Malo probably imagine that Saint-Malo is a very nice beach and historic city in France and very sunny. In fact, Saint-Malo can be very rainy, even in late June. So my favorite ISCA-related memory is the bright sun that was our companion at ISCA 2010."

## Muhammad Shahbaz

Favorite Memory:
“First ISCA paper: PMNet: In-Network Data Persistence”

## Tim Sherwood

Favorite Memory:
“In 1997 I was a plucky undergraduate at UC Davis with an interest in bioinformatics. A junior faculty member mentioned in their class that they were looking to hire a student for the summer and, as a kid in need of some money to help keep my student loans in check, I all of a sudden found myself quite out of my depth doing computer architecture for a Professor Fred Chong and his graduate student Mark Oskin. Setting a completely unrealistic expectation for the rest of my career, that work on how to virtualize in-memory computing led immediately to an ISCA paper in 1998 and in the process completely hooked me on the field of computer architecture. There are so many more people today interested in this wonderful discipline, and yet also so many more important ideas to discover. I appreciate all the hard work our community does to find and celebrate new contributions and I wish to offer a special congratulations to everyone having their first ISCA paper this year!”

## Jim Smith

Favorite Memory:
“When ISCA 1981 was held in Minneapolis, I was working for the Control Data Corporation just north of the Twin Cities. I was part of a team developing a deeply pipelined high performance processor and had devised a method for “conditional issue” (speculative execution) that relied on branch prediction. Because the conference was being held locally, management sent out a memo encouraging paper submissions (including a modest bonus). I wrote a paper on the branch prediction part of conditional issue, and it was accepted.

My research area when I went to CDC in 1979 was fault tolerant computing, so I was largely unfamiliar with the computer architecture research community. Within that community a “changing of the guard” was taking place, and the new guard was very open and welcoming to a newcomer. Up to that time, most of the innovations in high performance architecture had come from mainframe companies. However, by 1981 microprocessor technology had reached the point that high performance methods such as pipelining were becoming feasible. A new set of companies was entering the computer business, ready to leverage university research like never before. Excitement was in the air.

My clearest memory of the conference itself is Jim Thornton’s invited talk. Thornton was a hero of mine. He had been the lead designer of the CDC 6600 central processor, the chief architect for the Star-100 (a pioneering vector machine), and the founder of a pioneering LAN company that eventually lost out to Ethernet. He walked to the front of the auditorium, no podium, and gave a well-organized speech without the aid of notes. Old-school.

Having a paper at ISCA and the encouragement I got from other attendees gave me the confidence to change research areas, so in the Fall of 1981, I left CDC and returned to the University of Wisconsin, primed for a career in high performance computer architecture.

For the 8th ISCA, 90+ papers were submitted and 40 were accepted, 13 of which had a single author. The program committee was composed of 22 men. My submitted paper received two reviews, one of which pointed to two related papers and the other was a single sentence stating that the paper was good, but not excellent. The publications chair worked at IBM in Rochester and thought it would be a good idea if all the papers were in a "uniform and professional style". So someone in the publications department at IBM re-typed all of the papers, thereby producing an especially thick proceedings. I suggest you get a hard copy of the proceedings, leaf through all the pages for the tactile pleasure, and check out the state of computer architecture 42 years ago."

## Jakub Szefer

Favorite Memory:
"My first ISCA paper was in ISCA 2010: NoHype: Virtualized cloud infrastructure without virtualization. ISCA was in beautiful Saint-Malo, and our return flight got delayed by a day so I got to visit Paris during the extra night in France!"

## Josep Torrellas

Favorite Memory:
"The most memorable ISCA of all times must be Barcelona, ISCA 1998. The banquet at a palace in the evening was unbelievable, in terms of the food and the setting.

In addition, after an evening activity, I got to walk Maurice Wilkes to the hotel, and for over 30 min he told me about his pioneering experiences. What a treat!"

## David A. Wood

Favorite Memory:
"All of my favorite memories involve the informal social program events. Perhaps most notable was in Barcelona, when after the banquet I commandeered a bus to take John Hennessy, David Patterson, and Arvind (among a bus load of others) down to the disco/play area at the marina (Mare-something, I think) and playing miniature golf with them while disco music blasted us from all sides."

## Picture Gallery

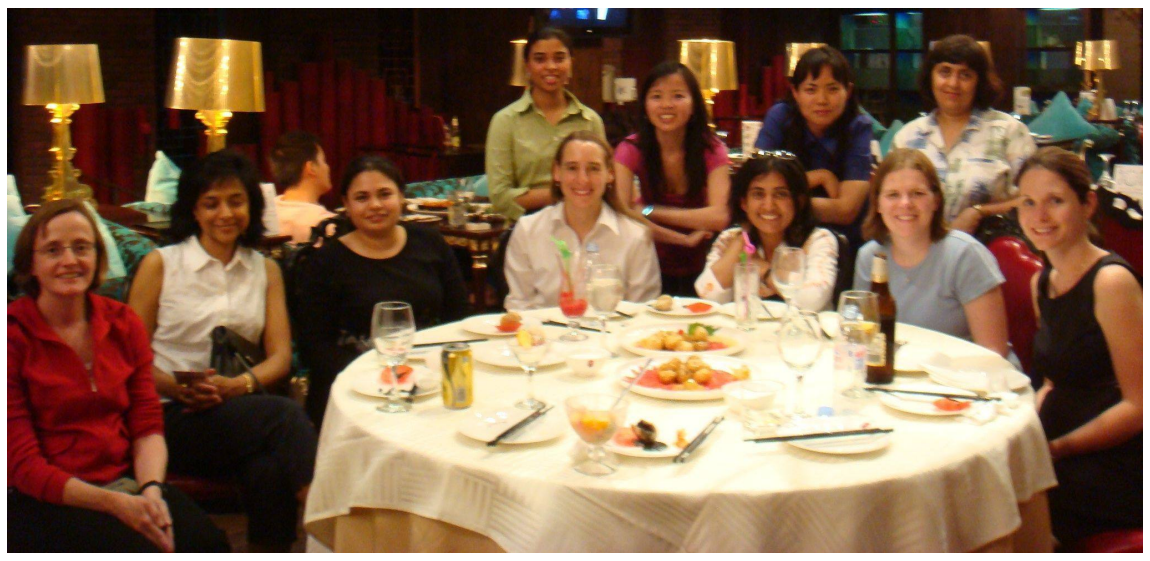

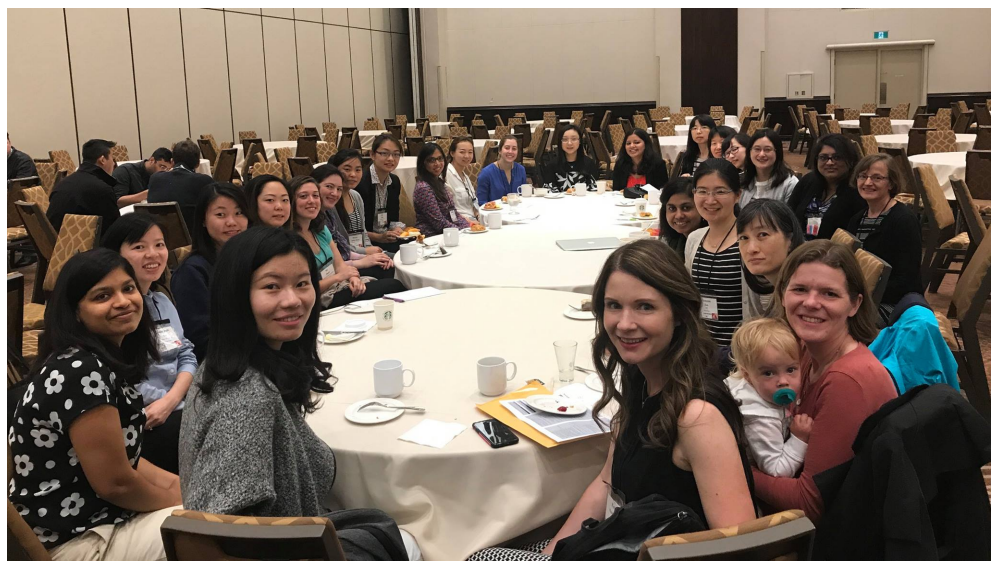

**-Carol-Jean Wu**

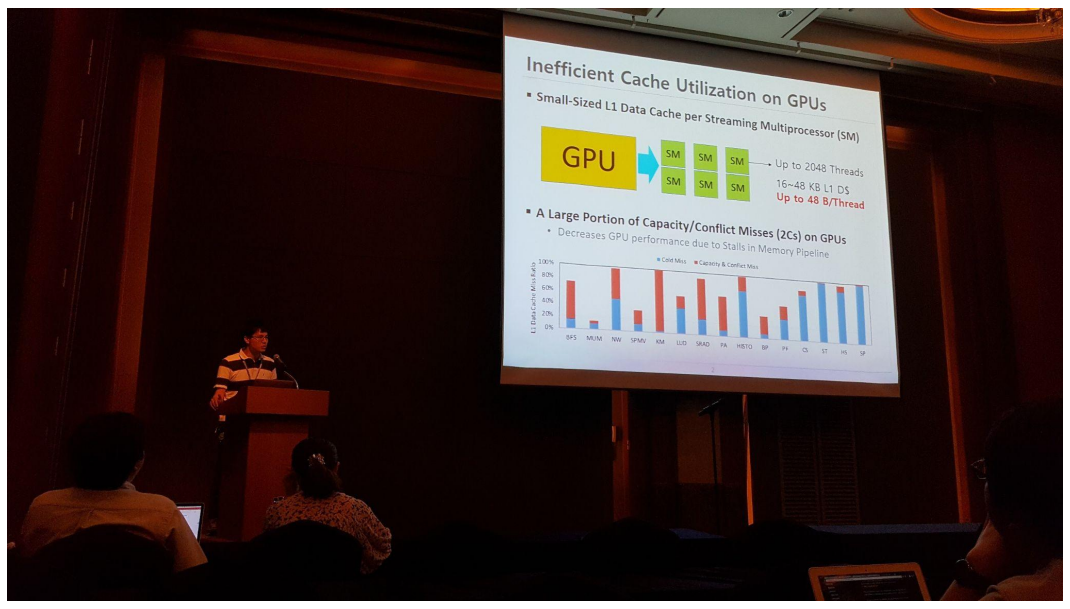

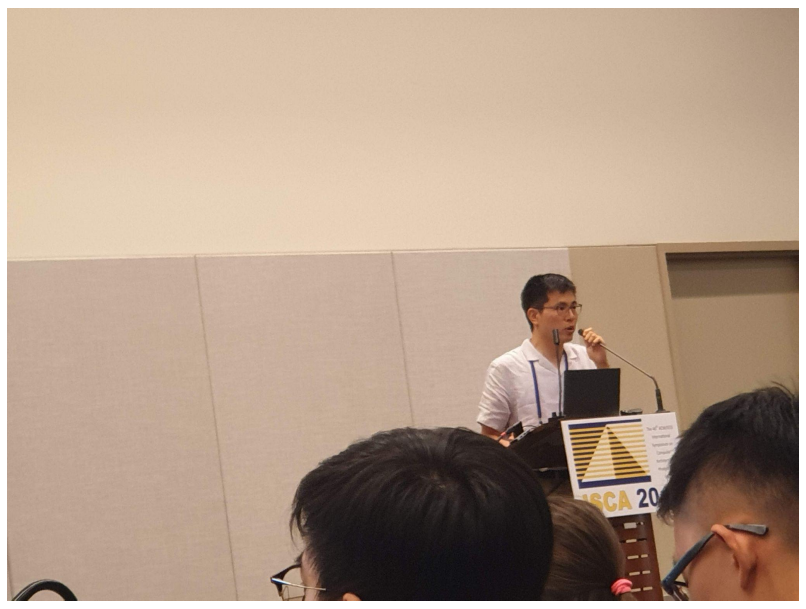

**- Yunho Oh**

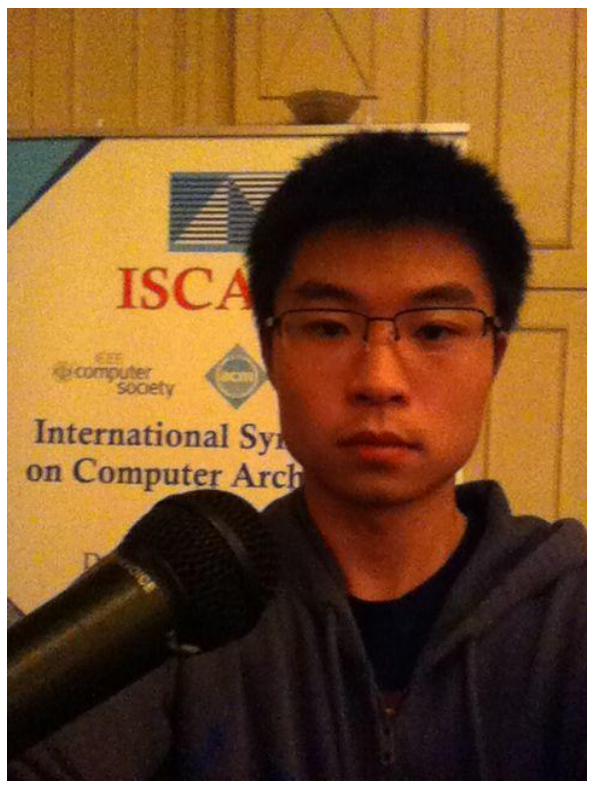

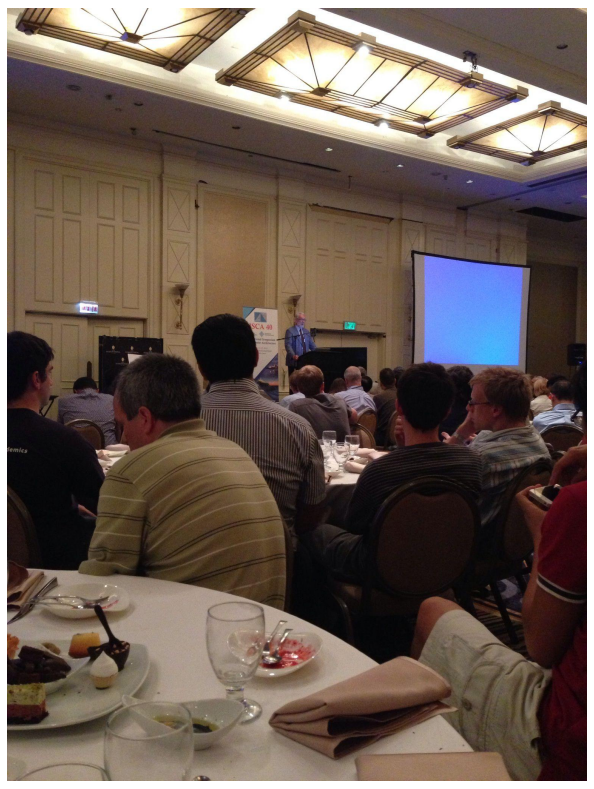

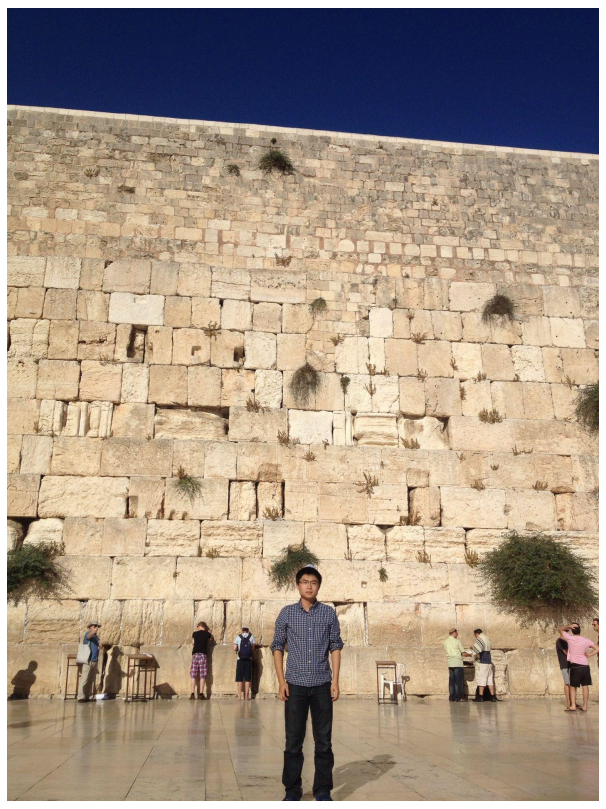

**- Jingwen Leng**

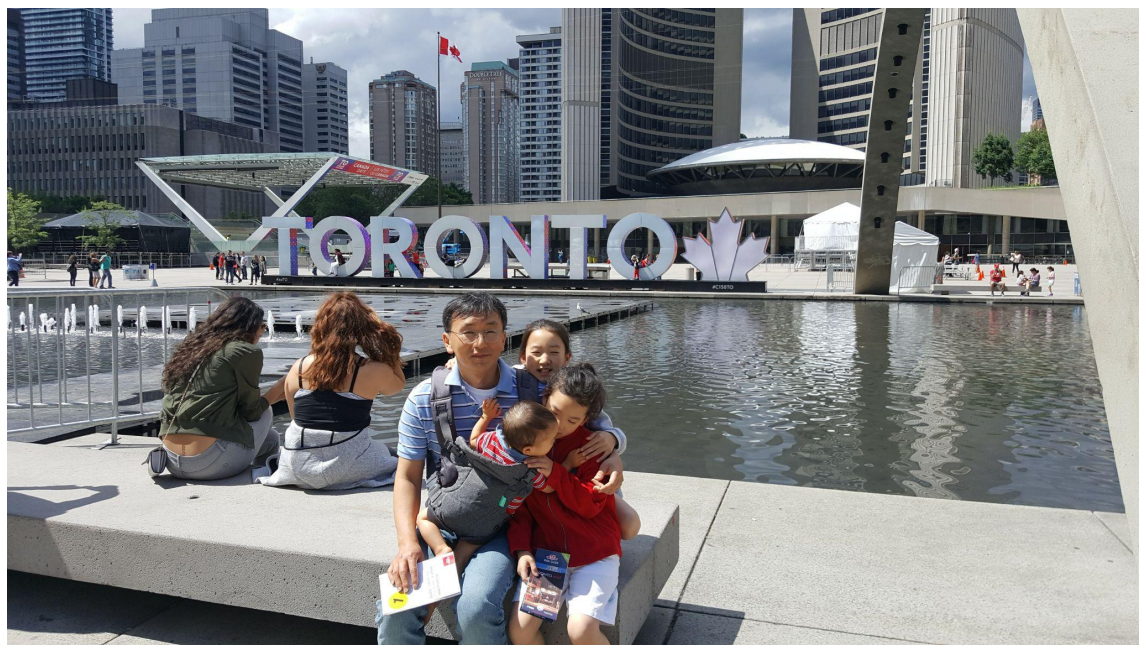

**- Nam Sung Kim**

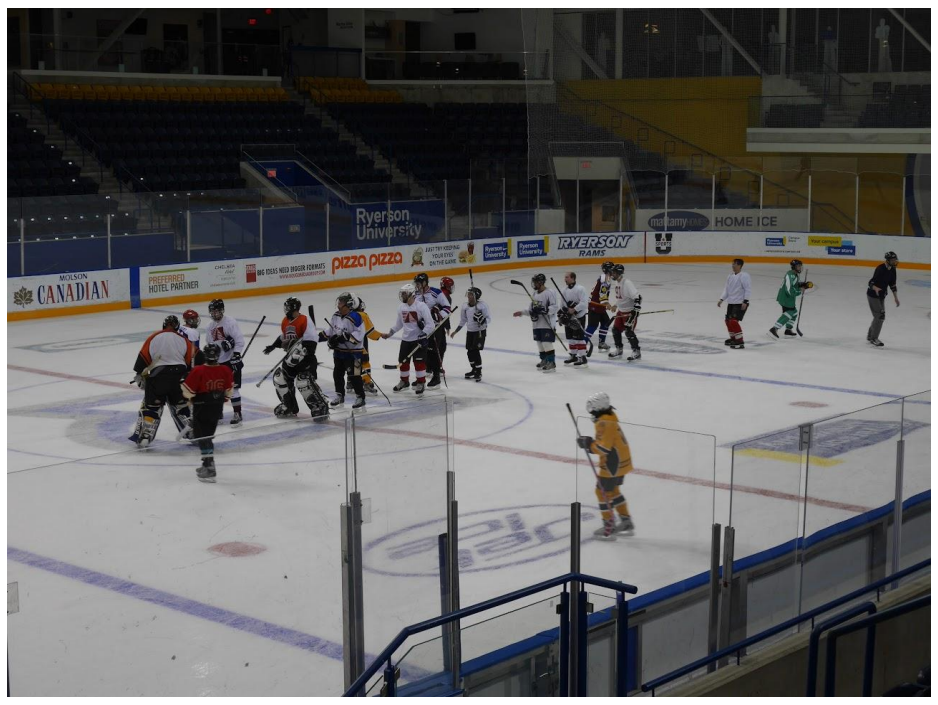

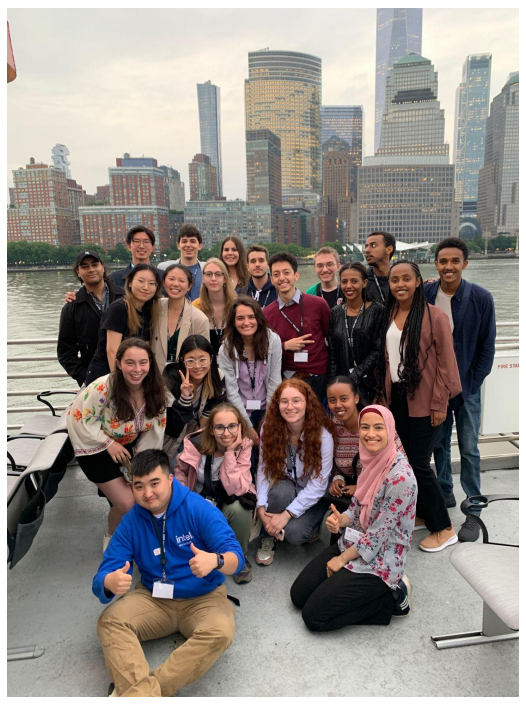

- **Joshua San Miguel (Photo credit: Brian Fu.)**

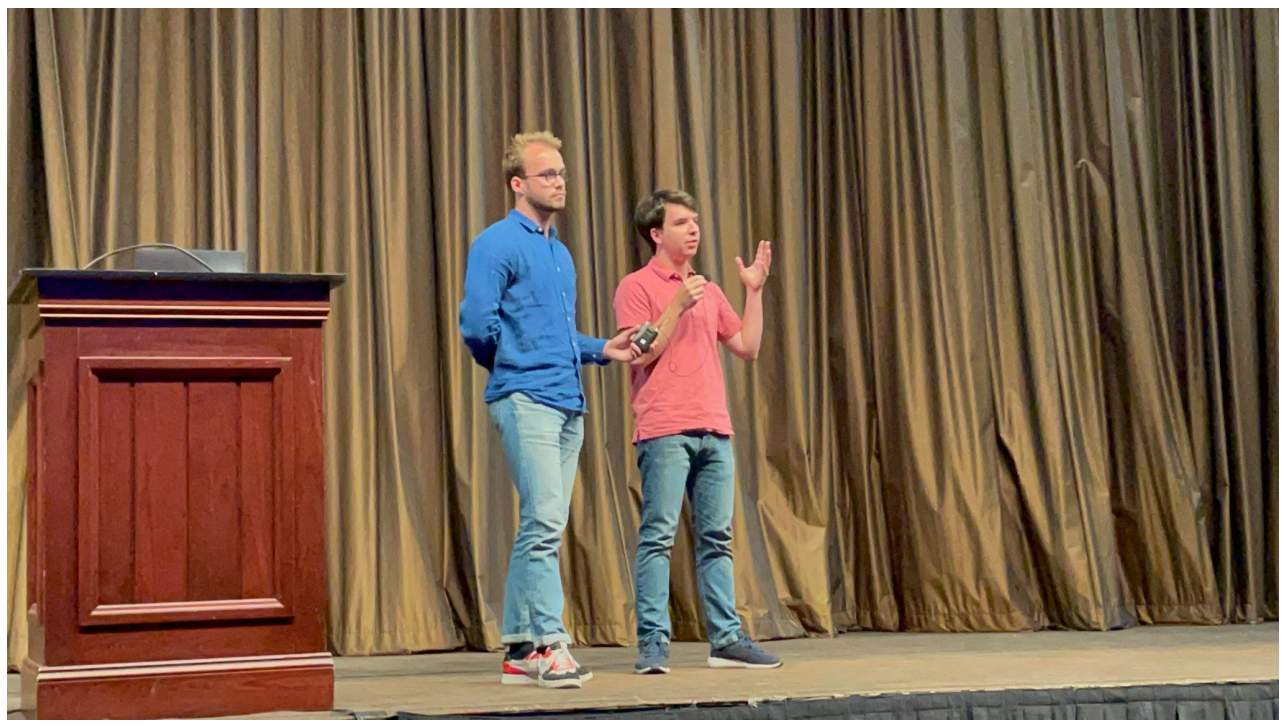

**- Jack Cook**

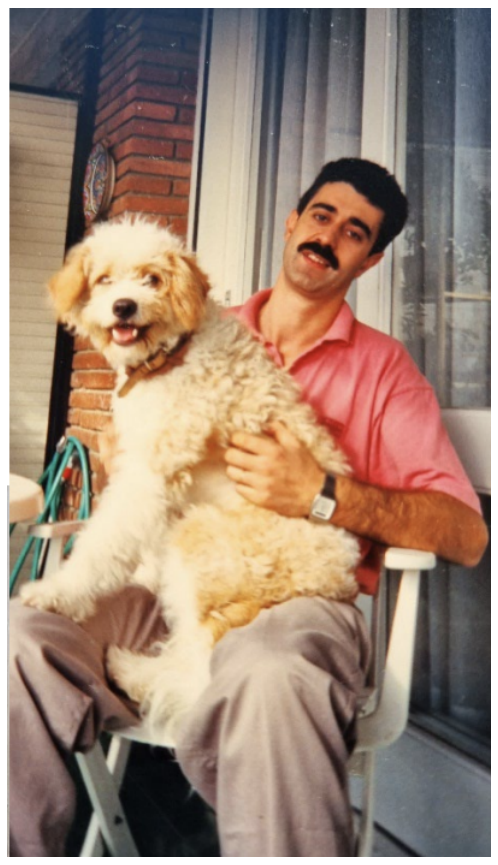

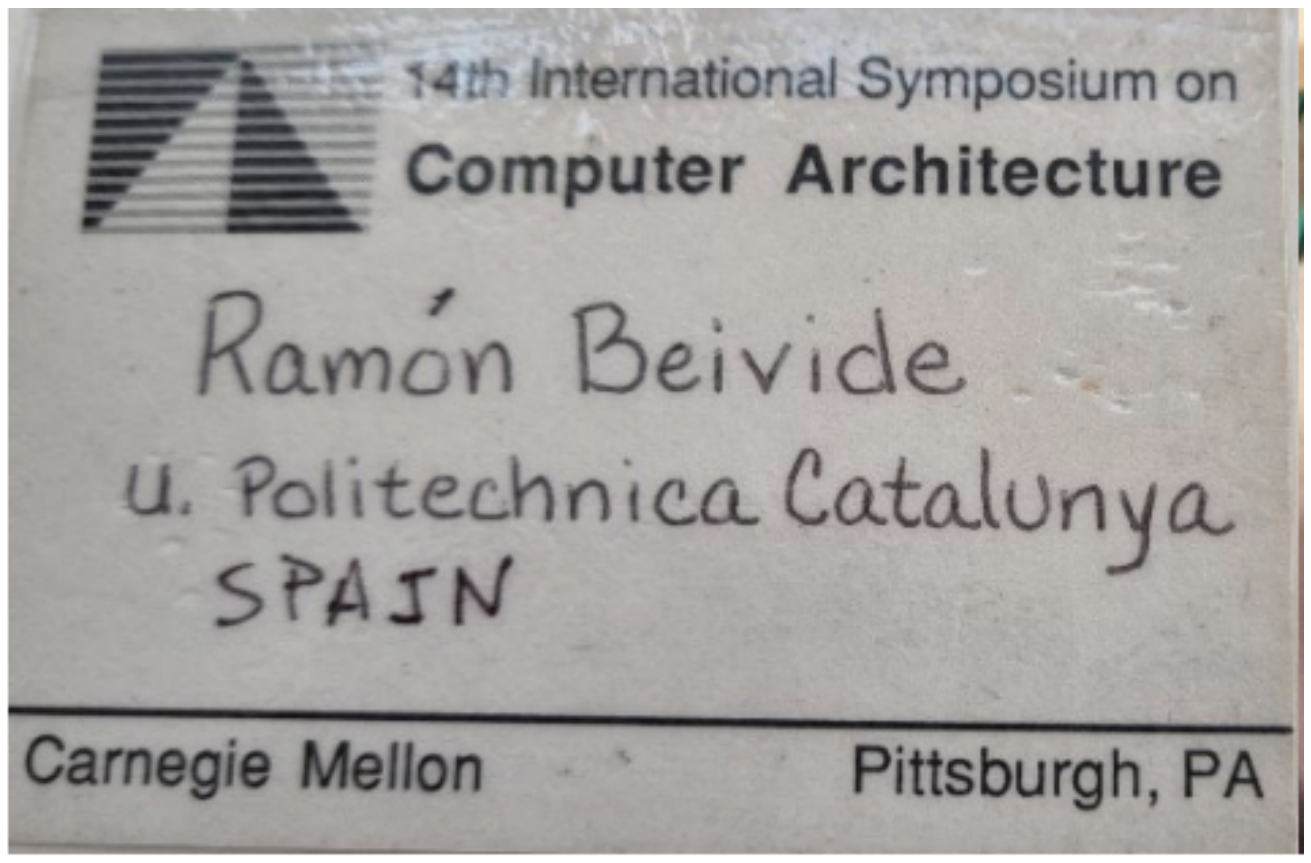

**- Ramon Belvide**

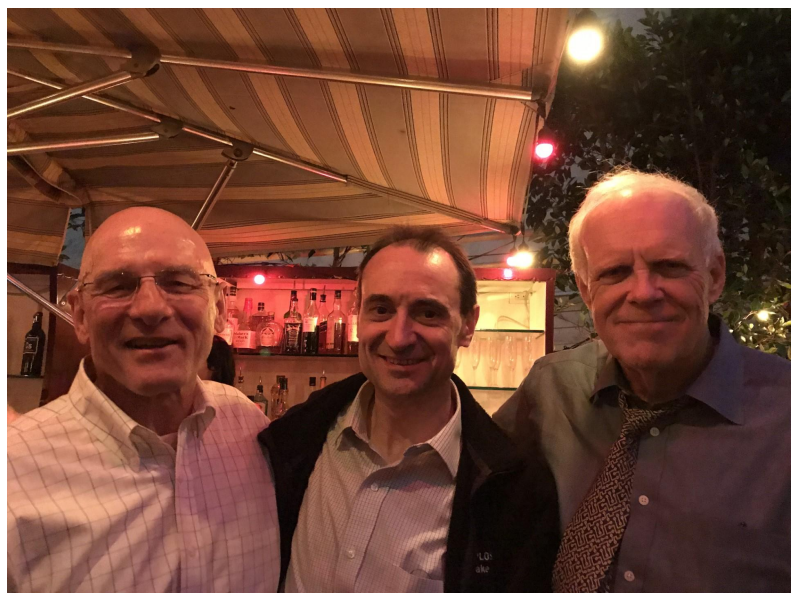

**- Josep Torrellas**